\documentclass[twocolumn]{emulateapj}  

\usepackage[left]{lineno}
\usepackage{epstopdf}
\usepackage{ulem}
\usepackage{amssymb}
\usepackage{natbib}
\usepackage{times}
\usepackage{graphicx}
\usepackage[usenames,dvipsnames]{pstricks}
\usepackage{psfrag}
\usepackage{lipsum}
\usepackage{arydshln}
\usepackage[colorlinks=true,linkcolor=blue,citecolor=blue]{hyperref}

\newcommand{\be}{\begin{equation}}
\newcommand{\ee}{\end{equation}}
\newcommand{\bary}{\begin{eqnarray}}
\newcommand{\eary}{\end{eqnarray}}

\shorttitle{GRB 190829A}

\begin{document}
\title{On the origin of the multi-GeV photons from the closest burst\\ with intermediate luminosity: GRB 190829A} 
\author{N.~ Fraija\altaffilmark{1$\dagger$}, P. Veres\altaffilmark{2},   P. Beniamini\altaffilmark{3},  A.~ Galvan-Gamez\altaffilmark{1},  B.~ D.~ Metzger\altaffilmark{4,5},  R. Barniol Duran\altaffilmark{6} and R. L. Becerra\altaffilmark{1}}
\affil{$^1$Instituto de Astronom\'ia, Universidad Nacional Aut\'{o}noma de M\'{e}xico, Apdo. Postal 70-264, Cd. Universitaria, Ciudad de M\'{e}xico 04510}
\affil{$^2$Center for Space Plasma and Aeronomic Research (CSPAR), University of Alabama in Huntsville, Huntsville, AL 35899, USA}
\affil{$^3$  TAPIR, Mailcode 350-17, California Institute of Technology, Pasadena, CA 91125, USA}
\affil{$^4$  Columbia Astrophysics Laboratory, Columbia University, New York, NY10027, USA}
\affil{$^5$  Center for Computational Astrophysics, Flatiron Institute, New York, NY 10010, USA}
\affil{$^6$  Department of Physics and Astronomy, California State University, Sacramento, 6000 J Street, Sacramento, CA 95819-6041, USA}
\email[$\dagger$ ]{nifraija@astro.unam.mx}
\begin{abstract}
Very-high-energy (VHE) emission is usually interpreted in the  synchrotron-self Compton (SSC) scenario, and expected from the low-redshift and high-luminosity gamma-ray bursts (GRBs), as  GRB 180720B and GRB 190114C.   Recently, VHE emission was detected  by the H.E.S.S. telescopes from one of the closest burst GRB 190829A which was associated with the supernova (SN) 2019oyw.   In this paper, we present a temporal and spectral analysis from optical bands to Fermi LAT energy range  over multiple observational periods beginning just after the BAT trigger time and extending for almost three months.      We show that the X-ray and optical observations are consistent with synchrotron forward-shock emission evolving between the characteristic and cooling spectral breaks during the early and late afterglow  in a uniform-density medium.   Modelling the light curves together with its spectral energy distribution, it is shown that the outflow expands with an initial bulk Lorentz factor of $\Gamma\sim 30$, which is high for a low-luminosity GRBs and low for a high-luminosity GRBs.     The values  of the initial bulk Lorentz factor and the isotropic equivalent energy suggest that GRB 190829A is classified as an intermediate-luminosity burst and consequently,  it becomes the first burst of this class in being detected  in the VHE gamma-ray band by an
imaging atmospheric Cherenkov telescope, and, in turn,  the first  event without being simultaneously observed by the Fermi LAT instrument.   Analyzing  the intermediate-luminosity bursts with $z\lesssim 0.2$ such as GRB 130702A, we show that bursts with intermediate luminosity are potential candidates to be detected in very-high energies.
\end{abstract}
%

\section{Introduction}
%
Observational evidence has firmly established that gamma-ray bursts (GRBs) lasting longer than few seconds are associated to the core collapse (CC) of massive stars \citep{1993ApJ...405..273W} leading to  supernovae \citep[SNe;][]{1998Natur.395..670G, 1999Natur.401..453B, 2006ARA&A..44..507W}.   Based on the isotropic equivalent luminosity in the gamma-ray band  and the opening angle,  some authors have classified CCSNe  as low-luminosity (ll)GRBs with  $L_{\rm iso}\lesssim 10^{48.5}\,{\rm erg\,s^{-1}}$, intermediate-luminosity (il)GRBs with  $10^{48.5} \lesssim L_{\rm iso}\lesssim 10^{49.5}\,{\rm erg\,s^{-1}}$ and high-luminosity (hl)GRBs with $L_{\rm iso}\gtrsim 10^{49.5}\,{\rm erg\,s^{-1}}$ \citep{2011ApJ...739L..55B, 2013RSPTA.37120275H, 2017A&A...605A.107C}.    Whereas  llGRBs  are associated with the shock break-out\footnote{If llGRBs are shock breakouts, then one should distinguish between the injected luminosity and the observed one, since they will be very different.} and characterized by having opening angles $\gtrsim 30^\circ$, and hlGRBs with an emerging collimated jet  \citep{2011ApJ...739L..55B}, there is no clear association for ilGRBs \citep{2014A&A...566A.102S}. To date, there are eight confirmed GRB-SNe detected within $z\lesssim 0.2$:  GRB 980425/ SN 1998bw \citep{1998Natur.395..670G},  GRB 060218/ SN 2006aj \citep{2006Natur.442.1008C},  GRB 100316D/ SN 2010bh \citep{2011ApJ...740...41C} and GRB 171205A/ SN 2017iuk \citep{2019Natur.565..324I} classified as llGRBs,   GRB 130702A/ SN 2013dx \citep{2015A&A...577A.116D}  as  ilGRB and GRB 030329/SN 2003dh  \citep{2003Natur.423..847H} as hlGRB.\footnote{Due to the luminosity and opening angle, GRB 161219B/SN 2016jca cannot be classified as low- or intermediate-lGRB \citep{2017A&A...605A.107C, 2019MNRAS.487.5824A}.}\\   
Very-high-energy (VHE $\geq$ 10 GeV) photons  are expected from low-redshift and high-luminosity bursts \citep{2019ApJ...884..117W, 2019ApJ...883..162F}.    During the afterglow phase electrons are shock-accelerated and cooled down by synchrotron process that radiates photons from radio to  gamma-rays.    The maximum photon energy radiated by the synchrotron process during the deceleration phase becomes $\sim 5 - 10 ~{\rm GeV}~\left(\frac{\Gamma(t)}{100}\right)\left(1+z\right)^{-1}$, where $\Gamma(t)$ is the bulk Lorentz factor (decaying with time) and $z$ is the redshift  \citep{2010ApJ...718L..63P, 2009ApJ...706L.138A, 2011MNRAS.412..522B}.     Another cooling process is the synchrotron self-Compton (SSC) emission, synchrotron photons are scattered above tens of GeVs  by the same electron population \citep{2001ApJ...559..110Z, 2019ApJ...883..162F, 2015ApJ...804..105F, 2017ApJ...848...94F}.    This is the case of GRB 180720B and GRB 190114C which showed similarities to other hlGRBs detected by Fermi LAT \citep{Ajello_2019}. These bursts were detected by the Major Atmospheric Gamma Imaging Cherenkov \citep[MAGIC;][]{2019Natur.575..459A}  and  the High Energy Stereoscopic System \citep[H.E.S.S.;][]{2019Natur.575..464A} telescopes at energies above 100  GeV, respectively.   The  high-energy photons detected by Fermi LAT beyond the synchrotron limit together with the VHE photons reported by H.E.S.S. and MAGIC telescopes were interpreted in the SSC scenario during the afterglow phase \citep{2019ApJ...885...29F, 2019ApJ...883..162F, 2019arXiv191109862Z}. On the other hand,  in the case of llGRBs which are characterized by being less energetic \citep{2011ApJ...739L..55B}, high-energy photons in the energy range of the Fermi LAT are hardly expected.   An interesting case study in this aspect is ilGRBs which have luminosities between low- and high-lGRBs \citep{2015A&A...577A.116D, 2017A&A...605A.107C}.\\
The Burst Area Telescope (BAT) instrument on-board the Swift satellite triggered on GRB 190829A at 2019 August 29 19:56:44.60 UT  \citep{2019GCN.25552....1D}.   The H.E.S.S. telescopes followed-up  the afterglow of GRB 190829A.   A preliminary onsite analysis of these observations showed a $>5 \sigma$  gamma-ray excess in coincidence with the direction of GRB 190829A \citep{2019GCN.25566....1D}.   This burst associated with a SN type Ic-BL \citep{2019GCN.25677....1D} was followed up by a large-scale campaign with several instruments onboard satellites and ground telescopes  that covered most of the electromagnetic spectrum.  Recently,  \cite{2020arXiv200100648C} discussed the VHE emission detected from GRB 190829A in terms of the  low-luminosity bursts and shock breakout scenario.\\
GRB 130702A classified as an intermediate-luminosity burst and associated to a broad-line, type Ic supernovae SN2013dx \citep{2015A&A...577A.116D}, was detected at different wavelengths ranging from radio to high-energy gamma-rays. The Gamma-ray Burst Monitor (GBM) instrument on-board the Fermi satellite triggered on GRB 130702A at 2013 July 02 00:05:23.079 UTC. The Fermi LAT instrument detected photons from this burst above $>$100 MeV  within 2200 s.  Details of the data analysis and the afterglow observations are reported in  \cite{2015A&A...577A.116D} and  \cite{2016ApJ...818...79T}.\\
In this paper,  we present a detailed data analysis of the multi-wavelength  observations of GRB 190829A.  Using the best-fit parameters found after modelling the X-ray and optical light curves of this burst, we analyze the VHE emission reported by the H.E.S.S. experiment.   Our model is generalized to study the mechanism involved to interpret the high-energy photons around other ilGRB (e.g. GRB 130207A).    The paper is arranged as follows. In Section 2 we present multi-wavelength observations and data reduction. In Section 3 we model and interpret the multi-wavelength observations. In Section 4 we discuss our results around GRB 190829A. In Section 5 we present the analysis and discussion of the multi-GeV photons reported in GRB 190829A and GRB 130702A and finally,  in Section 5 we give a brief summary.  The convention $Q_{\rm x}=Q/10^{\rm x}$  in cgs units will be adopted through this paper.  
%
%
%
\section{GRB 190829A}\label{sec:observations}
\subsection{Observations and  data reduction}

\subsubsection{Fermi: GBM observations}
The Fermi GBM instrument triggered and localized GRB 190829A at 2019 August 29  19:55:53.13 UT.    GBM data is analyzed using the public database at the Fermi website.\footnote{http://fermi.gsfc.nasa.gov/ssc/data}   Flux values are derived using  the spectral analysis package Rmfit version 432.\footnote{https://fermi.gsfc.nasa.gov/ssc/data/analysis/rmfit/}  The time tagged event (TTE) files of the NaI detectors (5,6, 7 and 9) and the BGO detector B1 are used to analyze the GBM data.
Table \ref{table1:gbm_analysis} corresponds to the values of the spectral analysis of GRB 190829A using GBM data during the time interval [-2.0 ; 68.0] s.  During this interval,  an initial pulse followed by a brighter peak are observed:  the initial pulse  between [-2.0 ; 12.0] s and the brighter peak between [46.0 ; 68.0] s which correspond to total isotropic energies of $E_{\rm \gamma, iso}=(9.151\pm 0.504)\times 10^{49}$ erg and $(2.967\pm 0.032)\times 10^{50}$ erg, respectively, and peak energies  of ($67.88\pm 23.3$) keV and  ($11.47\pm 0.360$) keV, respectively.  These are calculated considering the energy range of 1 keV - 10 MeV.   This table shows the time interval (column one), low-energy (column 2) and high-energy (column three) spectral indexes of the Band function, the peak energy (column four), the isotropic-equivalent energy (fifth column) and the observed flux (last column).  Although the power law with exponential cutoff (CPL),  the blackbody (BB, Planck function) and the Band function \citep{1993ApJ...413..281B} are considered,  the time-resolved spectra best fit to the Band function. This model is preferred over both the CPL and the BB models.  To assess the quality of a spectral fit, the traditional $\chi^2$ statistics is used.   We want to emphasize that a thermal photospheric emission in the GBM data is not observed.\\
\subsubsection{Fermi: LAT observations}
The Fermi LAT instrument performed a search for high-energy emission on different time windows around the position of this burst and  upper limits with a 95\% confidence level in the 0.1 - 1 GeV energy range were derived \citep{2019GCN.25574....1P}.  Considering a PL function $\propto \epsilon_\gamma^{-\Gamma_{\rm LAT}}$ with a photon index of $\Gamma_{\rm LAT}=\beta_{\rm LAT} + 1 =  2.0$, the derived LAT upper limits were $5.3\times 10^{-10}$, $3.2\times 10^{-10}$,  $1.4\times 10^{-10}$ and   $1.8\times 10^{-10}\,{\rm erg\,cm^{-2}\, s^{-1}}$ for time windows of 0 - 1.1,  0 - 10, 10 - 30 and 15 - 30 ks, respectively.\\
\subsubsection{Swift: UVOT observations}
The Swift UVOT started settled observations of the field of view of this burst 106 s after the BAT trigger time \citep{2019GCN.25570....1O}.  This instrument detected an emission consistent with the afterglow in the V, B, White and U bands.  Analyzing  the afterglow spectrum,  \cite{2019GCN.25565....1V} found absorption lines of Ca, H and K doublet, identifying this burst with DSS galaxy  J025810.28-085719.2 at a redshift of $z = 0.0785 \pm 0.005$.     The UVOT data is obtained using the  public available database at the official Swift website.\footnote{https://www.swift.ac.uk/archive/obs.php}  The  observed fluxes and their corresponding uncertainties used in this work are calculated using the standard conversion for AB magnitudes shown in \cite{1996AJ....111.1748F}.  The optical data is corrected by the galactic extinction using the relation derived in \cite{2019ApJ...872..118B}. The values of $\beta_{\rm O}=0.48$ for optical filters and a reddening of $E_{B-V}=0.05$ \citep{2019GCN.25552....1D} are used.\\
\vspace{1cm}
\subsubsection{Swift: BAT and XRT observations}
The Swift BAT instrument triggered on GRB 190829A at 2019 August 29 19:56:44.60 UT  \citep{2019GCN.25552....1D}.  The instrument located this burst with coordinates RA(J2000)=02h 58m 10s and Dec=-08d 58' 03'' with an uncertainty of 3 arcmin.  The Swift XRT instrument  started  detecting GRB 190829A at 19:58:21.9 UT, 97.3 s after the BAT trigger. This instrument monitored GRB 190829A in the Windowed Timing (WT) mode with a spectrum exposure of 128 s and the  Photon Counting (PC) mode with a spectrum exposure of 10.8 ks. Swift data is obtained using the  publicly available database at the official Swift website.\footnote{{\rm https://www.swift.ac.uk/burst\_analyser/00922968/}} The flux density at 10 keV is transformed to 1 keV using the conversion factor derived in \cite{2010A&A...519A.102E}.\\
\subsubsection{H.E.S.S.: VHE gamma-ray observations}
H.E.S.S. telescopes reported the detection of  VHE gamma-ray emission compatible with the direction of GRB 190829A \citep{2019GCN.25566....1D}.  This emission lasting 3.9 hours was detected  with a significance $\>$5$\sigma$ at 4.3 hours after the BAT trigger time. 
\subsubsection{GRB / SN observations}
GROND, mounted at the 2.2m MPG telescope at ESO La Silla Observatory,  found a relatively sharp growth in all bands between 4.5 and 5.5 days after the trigger time \citep{2019GCN.25651....1B}.  \cite{2019GCN.25664....1T}  studied  the spectrum of the optical afterglow with the Low Resolution Imaging Spectrometer (LRIS).  Authors found  identical features consistent with a broad-line SN and with the X-ray analysis reported by \cite{2019GCN.25568....1O}. In addition,  \cite{2019GCN.25677....1D} found evidence of broad absorption lines with expansion velocities similar to SN 1998bw. It confirmed the association of the SN 2019oyw classified as Type Ic-BL\footnote{http://www.rochesterastronomy.org/snimages/} with GRB 190829A.  
\subsection{Analysis of the multi-wavelength observations}
\subsubsection{GBM data analysis}
The upper left-hand panel in Figure \ref{figure1} shows the gamma-ray light curve and the evolution of the peak energy ($E_{\rm peak}$),  the low-energy ($\alpha_{\rm Band}$) and high-energy ($\beta_{\rm Band}$) spectral indexes of GRB 190829A.  The initial gamma-ray pulse is shown in open circles and  the brighter peak in filled circles. These sub-panels are ordered  from top to bottom: the low-energy spectral index, the high-energy spectral index, the energy peak and the gamma-ray light curve obtained in the 10 - 1000 keV energy range.    In order to fit the entire gamma-ray light curve, we use the functions given by  $F_\nu(t)\propto e^{-\frac{\tau_1}{t-t_0}}\,F_{\rm k}$ with $F_{\rm k}=e^{-\frac{t-t_0}{\tau_2}}$ \citep{2005ApJ...627..324N} for the initial pulse and  $F_{\rm k} =(\frac{t-t_0}{t_0})^{-\alpha_\gamma}$ \citep{2006Natur.442..172V} for the brighter peak. The terms $t_0$ is the starting time, $\tau_{1(2)}$ is the timescale of the flux rise (decay) and $\alpha_\gamma$  the power index of fast decay.  To fit the evolution of the parameters ($E_{\rm peak}$, $\alpha_{\rm Band}$ and $\beta_{\rm Band}$),  we use PL functions $\propto t^{-\delta}$ with  $\delta$ the power index.   The Chi-square ($\chi^2$) minimization method, developed in the ROOT software package \citep{1997NIMPA.389...81B} is used to find the best-fit values  which are reported in Table \ref{table2:fit_GBM}.  In order to find the minimum variability timescale for this burst,  the method utilized by \cite{2012ApJ...744..141B} is used.  This variability timescale corresponds to ($214.7 \pm 6.9$) ms.\\ 

\paragraph{Evolution of the spectral parameters}\label{evol}
During the first pulse, we found that the lower-energy photon index is around $\alpha_{\rm Band}\approx -0.1$ which  is much harder than the typical value ($\alpha_{\rm Band }\approx -1$) and also beyond the so-called  ``synchrotron line of death''  \citep[$-\frac23$;][]{1998ApJ...506L..23P}.  While  the spectrum parameters might indicate a contribution of thermal emission \citep{2018NatAs...2...69Z, 2010ApJ...709L.172R, 2000ApJ...530..292M},  magnetic dissipation \citep{2014MNRAS.445.3892B, 2017MNRAS.472.3058B, 2018MNRAS.476.1785B} or ``jitter" (a variant of the synchrotron)  radiation mechanism \citep{2000ApJ...540..704M},   the lower-energy photon index here is not well constrained, because the peak energy is very low, and the photons that were supposed to constrain $\alpha_{\rm Band}$ fall out of the GBM range. The scenarios of the jitter radiation and thermal emission are disfavored. The jitter due to the assumed small coherence scale ($\lambda_B$) is not revealed from particle-in-cell (PIC) numerical simulations of relativistic shocks \citep{2008ApJ...682L...5S,2008ApJ...673L..39S}  and the thermal emission due to the rapid variation of  the energy peak \citep{2009ApJ...702.1211R}.\footnote{Authors studied the evolution of the temperature in 49 individual time-resolved pulses.  They reported that during the first seconds the temperature is almost constant and after decay as a   PL with a sample-averaged temporal index of $0.68$.}\\
During the brighter gamma-ray peak the Band-function parameters are typical and similar to most bursts ($\alpha_{\rm Band}\approx -1.1$ and $\beta_{\rm Band}\approx -2.5$).   The value of low-energy spectral index of $\sim - 1$  can be interpreted as synchrotron emission in a decay magnetic field \citep{2014NatPh..10..351U}, magnetic reconnections/dissipations \citep{2014MNRAS.445.3892B, 2017MNRAS.472.3058B, 2018MNRAS.476.1785B, 2019ApJ...882..184L} and dissipative photosphere models \citep{2006ApJ...652..482P, 2010ApJ...725.1137L, 2016ApJ...831..175V, 2015MNRAS.454L..31A}.  It is also possible to have a combination of synchrotron and photospheric components that evolve differently with time \citep[e.g.][]{2017MNRAS.468.3202B}.\\
%
%
\subsubsection{Fermi data Analysis}
The upper right-hand panel in Figure \ref{figure1} shows the upper limits and the VHE emission in the range reported by H.E.S.S.  In order to obtain high-energy photons with energies above $\geq 100$ MeV, we analyze the data files of GRB 190829A which are given at the data website.\footnote{https://fermi.gsfc.nasa.gov/cgi-bin/ssc/LAT/LATDataQuery.cgi} Details of the Fermi tools and the procedure performed to analyze the Fermi LAT data are presented in \cite{2019ApJ...885...29F}. Our analysis indicates that there are no photons associated to GRB 190829A with a probability larger than 95\%.
\subsubsection{UVOT data analysis}
The lower left-hand panel in Figure \ref{figure1} shows the UV/optical light curves of GRB 190829A with the best-fit curves using broken power-law (BPL) functions. The light curves of the Swift UVOT obtained in the V, B, White and U bands display that before $\approx 700\,{\rm s}$ all color filters remain constant, after they increase as  $F_{\nu}\propto t^{-\alpha_{\rm O}}$ with  $\alpha_{\rm O} \approx -3$,  reaching the maximum values  at $\approx 1.4\times 10^3\,{\rm s}$, and finally these begin to decrease with slopes of $\alpha_{\rm O} \approx 1$ up to $\sim 3\times 10^4\,{\rm s}$. After all this time,  the flux densities in all color filters  become constant,  dominated by the host galaxy.  The corresponding best-fit values of each color filter as well as the timescales of $\Delta t/t$ \citep{2007ApJ...671.1903C} are reported in Table \ref{table4:optical}.   The timescales of $\Delta t/t$ are calculated using  $\Delta t$ the full width at half-maximum \citep[FWHM;][]{2007ApJ...671.1903C} adopting a Gaussian width and $t$ obtained at the maximum value of the flux.\\ 
 \subsubsection{BAT and XRT data analysis}
\begin{figure}[h!]
{ \centering
\resizebox*{0.48\textwidth}{0.3\textheight}
{\includegraphics{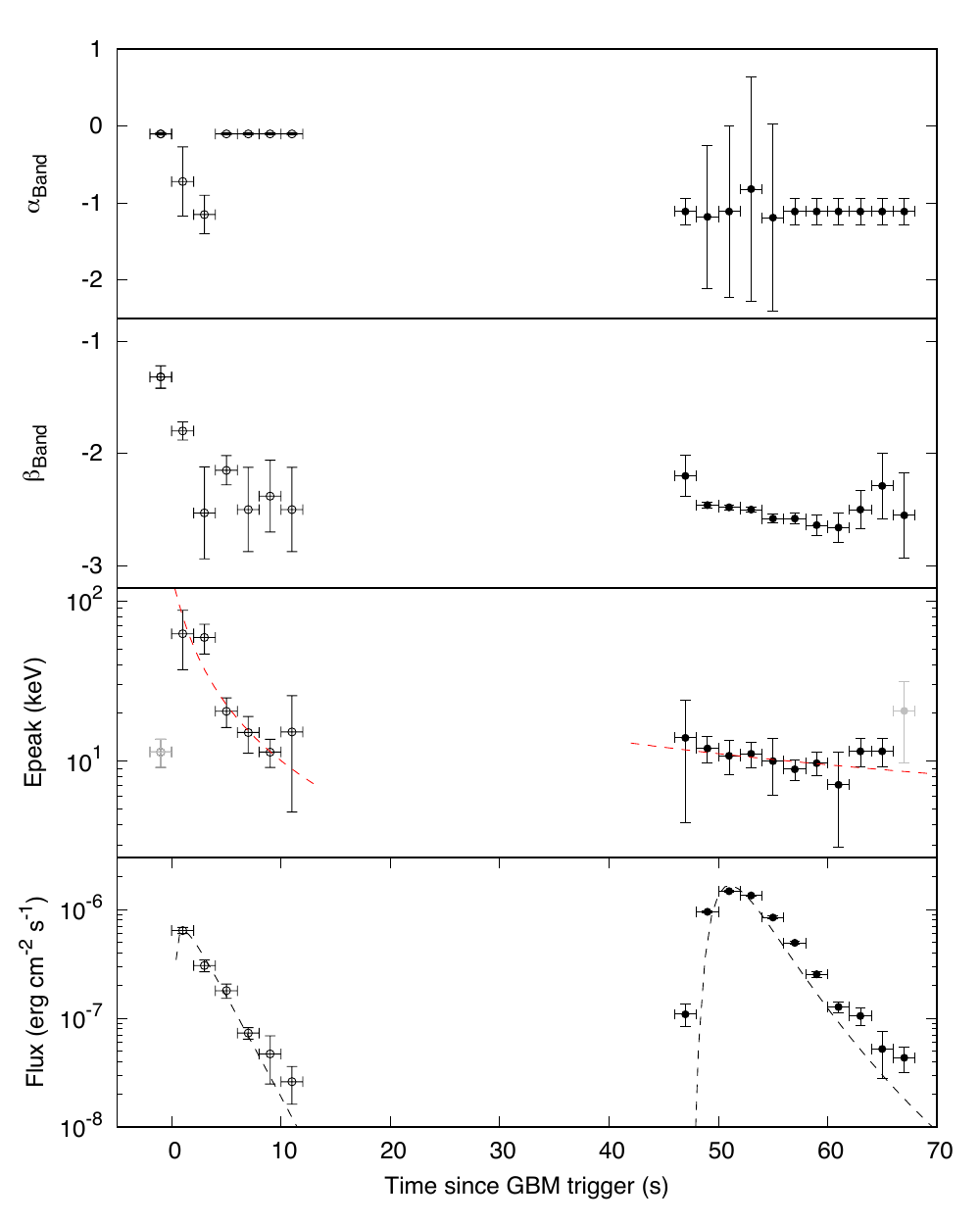}}
\resizebox*{0.48\textwidth}{0.3\textheight}
{\includegraphics{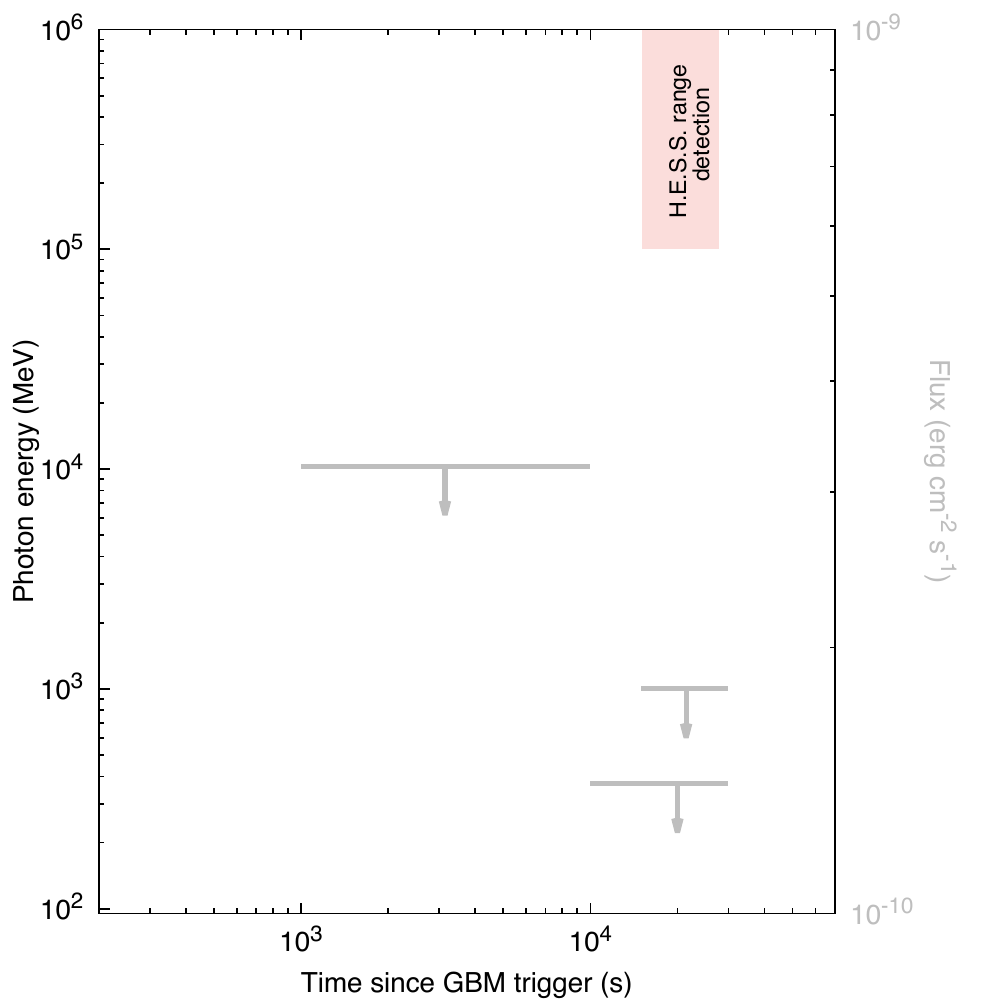}}
\resizebox*{0.48\textwidth}{0.3\textheight}
{\includegraphics{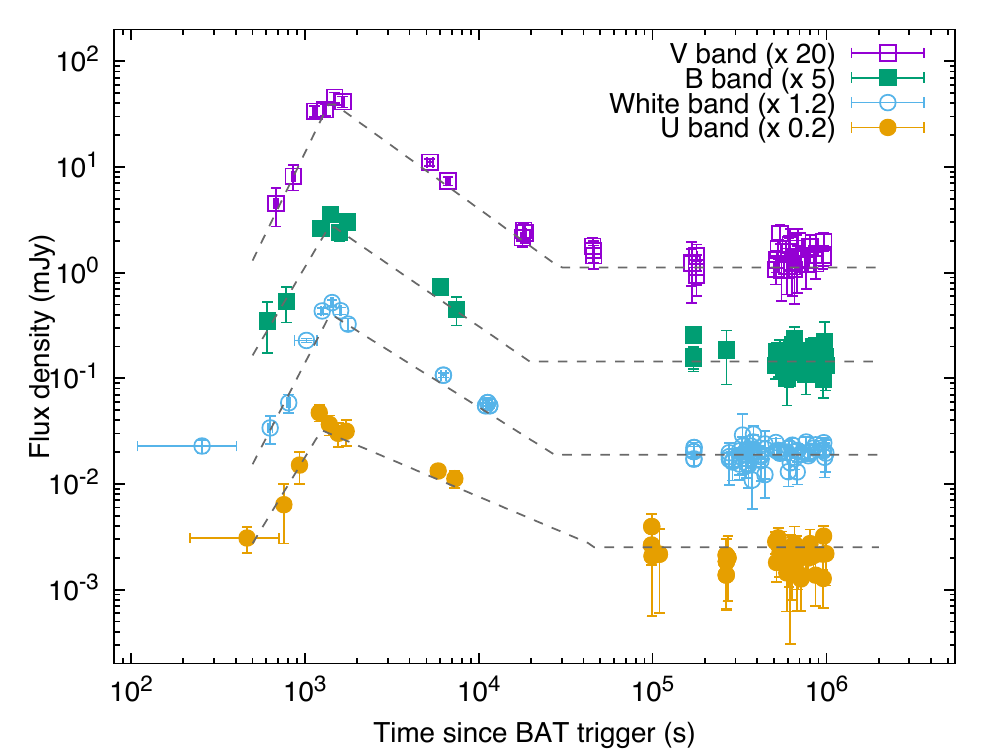}}
\resizebox*{0.49\textwidth}{0.3\textheight}
{\includegraphics{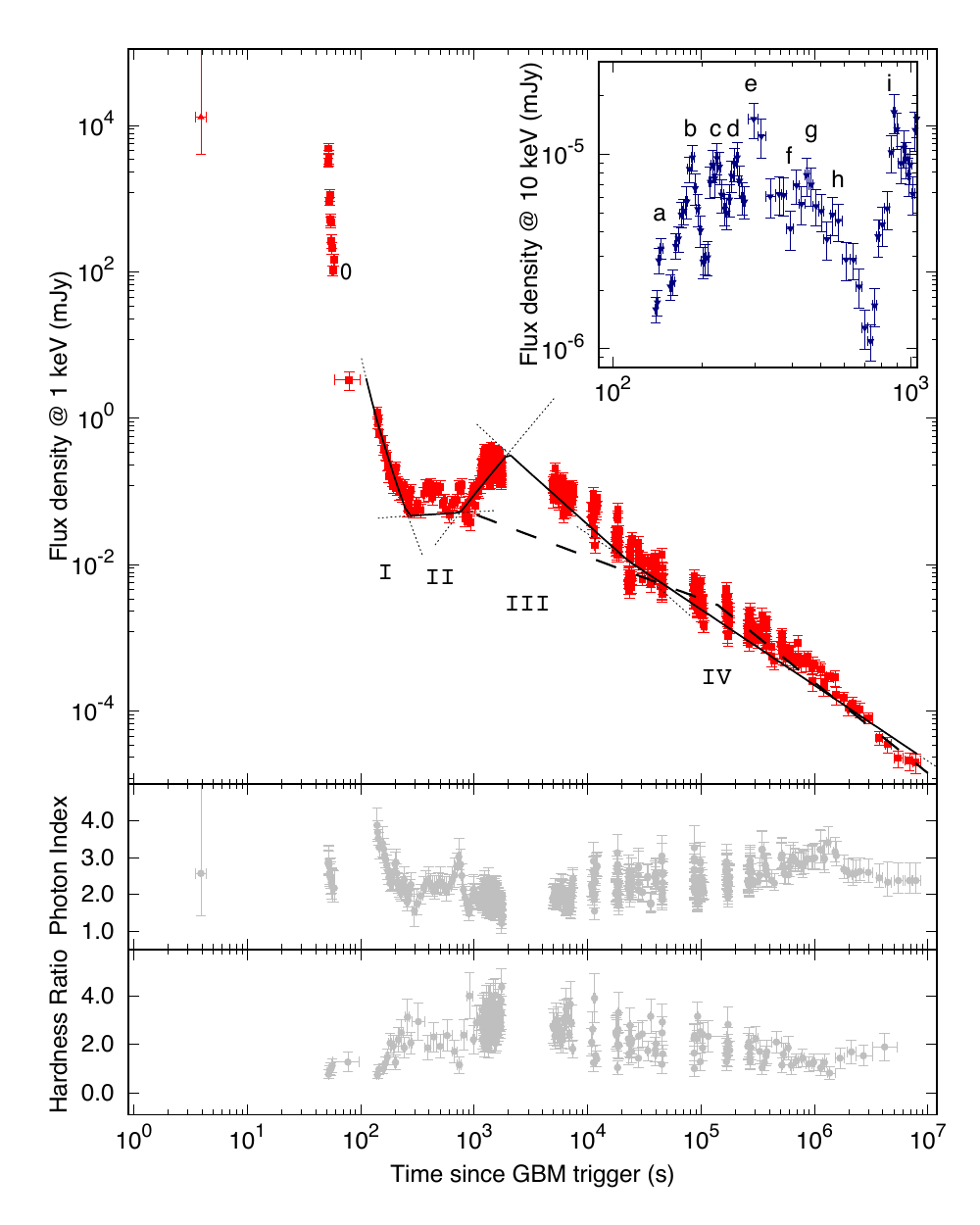}}
}
\caption{The upper left-hand panel shows the gamma-ray light curve and the evolution of the peak energy, the low-energy and high-energy spectral indexes of GRB 190829A. From top to bottom: the low-energy spectral index, the high-energy spectral index, the energy peak and the gamma-ray light curve obtained in the 10 - 1000 keV energy range. The initial gamma-ray pulse is shown in open circles and  the brighter peak in filled circles. Lines in all panels correspond to the best-fit functions.  Fermi GBM data  are reduced using the public database at their respective websites.  The upper right-hand panel shows Fermi LAT upper limits and all the photons with energies above 100 MeV with different probabilities of being associated to GRB 190829A. The lower left-hand panel shows the Swift UVOT light curve in the  V,  B, White and U bands.  The dashed grey curves in all color filters correspond to the best-fit BPL functions.   Swift UVOT data  are reduced using the public database at the official Swift web site. The lower right-hand panel shows the X-ray light curve at 1 keV and the small box at 10 keV.  At 1 keV five phases are labeled:  ``0" the initial PL segment , ``I"  the steep decay , ``II"   the plateau phase, ``III" X-ray flare and ``IV" the canonical normal decay and at 10 keV nine small peaks are labeled: ``a", ``b", ``c", ``d", ``e", ``f", ``g", ``h" and ``i".  The best-fit curves are drawn in each phase.  The middle sub-panel corresponds to the photon index light curves and the bottom sub-panel the hardness ratio light curve. The solid lines in black  corresponds to the best-fit curves found in this work and  the dashed lines in gray  represent the best-fit curves reported by Swift team.}

\label{figure1}
\end{figure}

\begin{figure}[h!]
{\centering
\resizebox*{0.5\textwidth}{0.3\textheight}
{\includegraphics{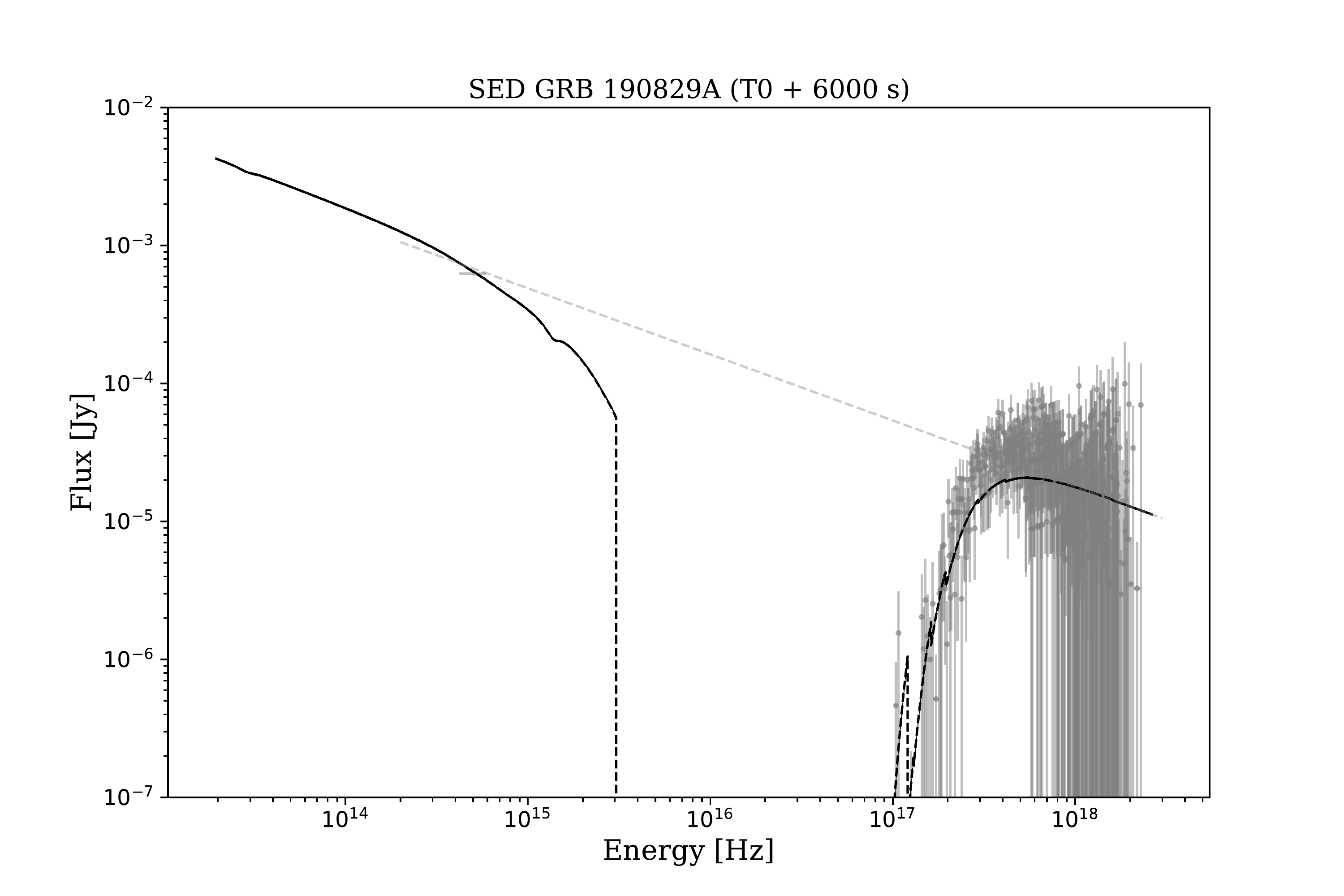}}
\resizebox*{0.5\textwidth}{0.3\textheight}
{\includegraphics{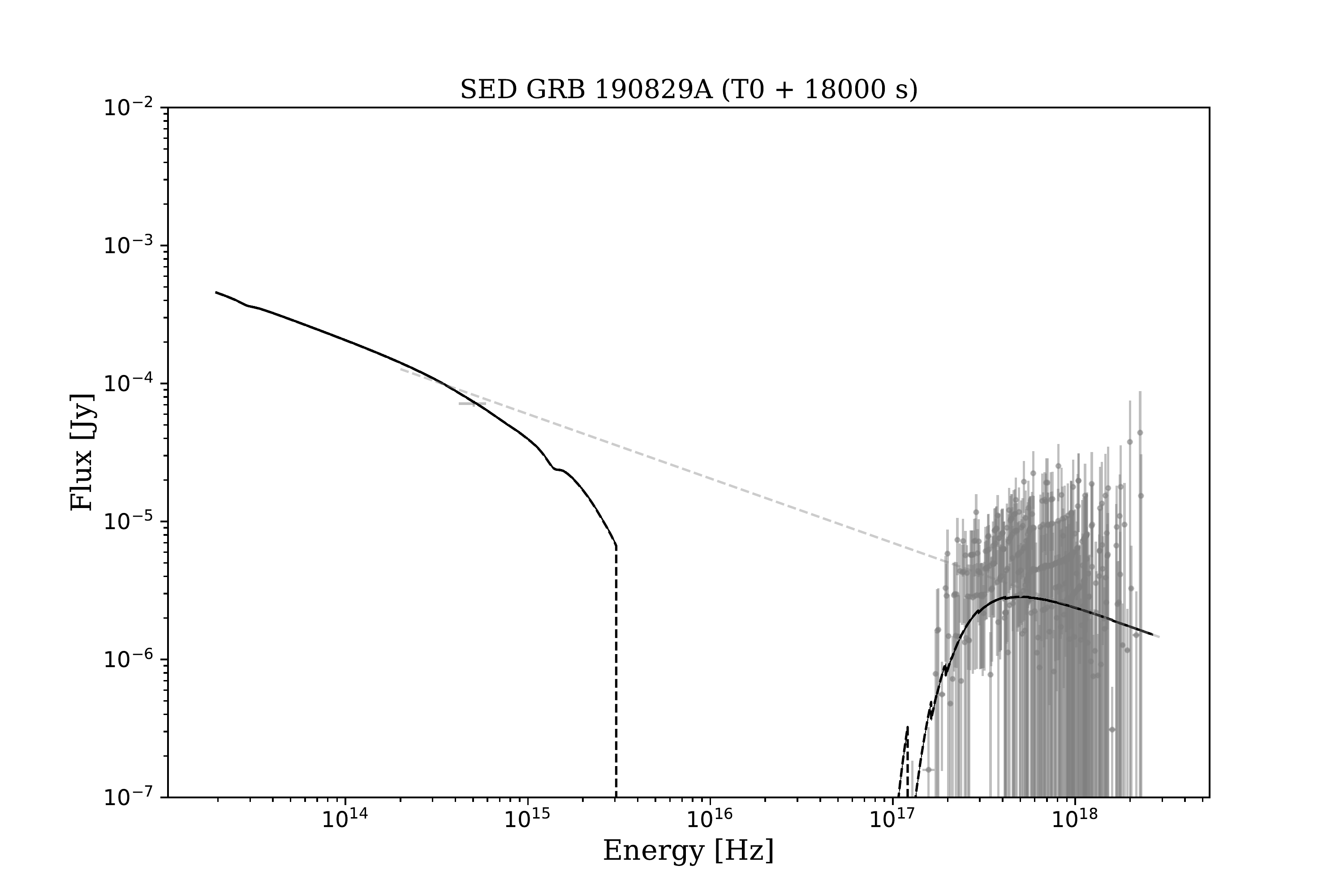}}
}
\caption{The broadband SEDs of the X-ray and optical observations at 6000 s (left) and 18000 s (right ) are shown (epoch ``III" ). The dashed gray lines in each panel correspond to the best-fit curve from XSPEC.}
\label{figure2}
\end{figure}

The lower right-hand panel in Figure \ref{figure1} shows the X-ray light curves (top), the spectral evolution of the photon index (medium) and hardness ratio (bottom). The X-ray light curve in the main panel is shown at 1 keV.
Five epochs, labeled as ``0", ``I", ``II", ``III" and ``IV" corresponding to the time intervals $[52 ; 62]$ s, $[132 ; 267]$ s, $[267 ; 752]$ s, $[752 ; 1.4\times10^5]$ s and  $> 1.4\times10^5$, respectively, are identified.\footnote{An additional epoch associated to an X-ray flare observed at 10 keV within a timescale of $\sim 10^6\,{\rm s}$ could be considered.}   A precursor can be identified in this panel $\sim$ 3 s after the GBM trigger time.  The dashed lines in gray color represent the best-fit curves reported by Swift team.\footnote{https://www.swift.ac.uk/xrt\_live\_cat/00922968/}
All the epochs are fitted with PL functions $\propto t^{-\alpha}$.  The epoch ``0" corresponds to the prompt emission fitted with a temporal index of $\alpha_{\rm X,I}=2.73\pm0.28$.   The steep decay in epoch ``I" is fitted with a temporal index of $\alpha_{\rm X,I}=3.53\pm0.70$ and  the subsequent very shallow decay identified in epoch ``II" is described with the temporal index of $\alpha_{\rm X,II}=0.06\pm0.05$.   In epoch ``III" the X-ray flare is modelled with a rising slope of $\alpha_{\rm X,III,rise}=-(3.12\pm0.94)$,  corresponding to a timescale of $\Delta t/t=0.75\pm0.24$.  Finally,  the  canonical normal decay  in epoch ``IV" is described using the values reported by the Swift analysis;  a temporal index of $\alpha_{\rm X,IV}=1.23\pm0.04$  after  the break time of $t_{\rm br}=1.4^{+0.17}_{-0.15}\times 10^5\,{\rm s}$.  The best-fit values of the X-ray data are reported in Table \ref{table3:X_ray}. \\
During the epoch ``I" and ``II" (see the small box),  the X-ray light curve at 10 keV displays several peaks labelled as ``a", ``b", ``c", ``d", ``e", ``f", ``g", ``h" and ``i".   The best-fit values of  the rising and falling slopes with their corresponding timescales are reported in Table \ref{table4:peaks}. \\
Figure \ref{figure2} shows the spectral energy distribution (SED) of GRB 190829A built with the optical and X-ray data at two different times (epoch ``III" ); $6.0\times 10^3$ s (left) and $1.8\times 10^4$ s (right).  These SEDs are modelled with SPLs with spectral indexes of $\beta_{\rm X}=0.48\pm 0.05$ and $\beta_{\rm X}=0.47\pm 0.05$, respectively.  The gray dashed lines correspond to the best-fit curves obtained from XSPEC for a column density of $(7.9\pm 0.6)\times 10^{21}\,{\rm cm^{-2}}$.\\ 
On the other hand, we analyze XRT data in the PC and WT mode with a PL and a BB model incorporated in XSPEC  v.12.10.1 \citep{1996ASPC..101...17A} in 13 time windows between $100\, {\rm s}$ and $1700\, {\rm s}$, as shown in Table \ref{tab:my-table}. This table shows the best-fit values using the PL and  the PL  plus  BB model. It can be seen that although in some time windows the traditional $\chi^2$ statistics is slightly better when a BB model is introduced, there is no clear evidence  of the existence of a thermal component.
\paragraph{The photon index and hardness ratio light curves}\label{evolution}
The lower right-hand panel in Figure \ref{figure1}  shows the spectral index ($\Gamma_{\rm ph}$; medium) and hardness ratio ($H_{\rm R}$; bottom) light curves.  The hardness ratio for the BAT data is defined by the photon flux in the ranges  (25 - 50 keV)/(15 - 25 keV) and for XRT data in the ranges (1.5 - 10 keV)/(0.3 - 1.5 keV) \citep{2010A&A...519A.102E}. We analyze the evolution of the photon index and hardness ratio in each epoch.
\subparagraph{Epoch ``0".} During the prompt emission it can be noted that the photon index and the hardness ratio  exhibit  a small degree of evolution. The photon index  increases progressively from $\Gamma_{\rm ph, 0}=0.18$ to $ 0.28$, and the hardness ratio decreases from $H_{\rm R,0}= 2.89$ to $2.17$.  During this epoch, a soft-to-hard spectral evolution is displayed.   The Pearson's correlation coefficient between the X-ray flux and the photon index is $r_{\rm  F\cdot\Gamma_{\rm ph}}=0.95$ with a p-value of  $2\times 10^{-6}$. It indicates that both variables are strongly correlated.
\subparagraph{Epoch ``I".} During the steep decay, the photon index and the hardness ratio light curves exhibit the strongest evolution.  These parameters vary significantly from $\Gamma_{\rm ph, I}=4.15$ to $1.24$, and from $H_{\rm R,I}= 0.59$ to $4.66$, respectively.  The maximum and minimum photon index corresponds to the maximum and minimum values of the light curve.  Similarly, the maximum and minimum hardness ratio corresponds to the maximum and one of the minimum values of the light curve.   During this epoch it is shown a soft-to-hard spectral evolution.    Using the ROOT software package \citep{1997NIMPA.389...81B}, we successfully fit the photon index and hardness ratio as $\Gamma_{\rm ph,I}\propto t^{-0.75\pm0.06}$ ($\chi^2/ndf=0.96$) and $H_{\rm R,I} \propto t^{1.36\pm0.19}$ (0.95), respectively. The Pearson's correlation coefficient between the flux and the photon index is $r_{\rm  F\cdot \Gamma_{\rm ph}}=0.77$ with a p-value of  $3\times 10^{-7}$.   It indicates a strong correlation between both variables.
\subparagraph{Epoch ``II"} During this epoch, the photon index and the hardness ratio evolve rapidly among two maximum and minimum values.  First,  the photon index and the hardness ratio vary from $\Gamma_{\rm ph,II}=1.34$ to $2.32$, and from $H_{\rm R, II}= 3.05$ to $1.79$, respectively.    Once the X-ray flux at 1 keV and 10 keV begins decreasing and increasing, respectively,  the photon index and the hardness ratio have a small decrease and increase, respectively, and then keep constant around $\approx 2$. At the end of interval,  the photon index and the hardness ratio have a small  increase and decrease, respectively, and then these parameters evolve strongly from  $\Gamma_{\rm ph,II}=2.61$ to $1.31$, and from $H_{\rm R,II}= 4.34$ to $1.45$, respectively.
\subparagraph{Epoch ``III"} During the X-ray flare,  the photon index and the hardness ratio displayed random fluctuation around $\approx 3$ and  $\approx 2.5$, respectively.
\subparagraph{Epoch ``IV"} During the  canonical normal decay, a very moderate spectral softening is observed.   
\section{Interpretation and Modelling of the multi-wavelength observations}
\subsection{Light curves from energy injection by a millisecond magnetar}\label{magnetar}  
The energy reservoir of a millisecond magnetar is the total rotation energy which is given by 
\be\label{Erot}
E_{\rm}=\frac12 I\, \Omega^2\,\approx 2.6 \times 10^{52}\,{\rm erg}\,M^{\frac32}_{\rm ns,1.4}\,P^{-2}_{-3}\,,
\ee
where $P$ is the spin period associated to an angular frequency $\Omega=2\pi/P$ and $I\simeq 1.3\times 10^{45}\,M^{\frac32}_{\rm ns,1.4}\,{\rm g\,cm^2}$ \citep{2005ApJ...629..979L} is the NS moment of inertia  with $M_{\rm ns}=1.4\, M_\odot\, M_{\rm ns,1.4}$ the NS mass.\\
CC-SNe are usually expected to leave a fraction of the stellar progenitor bound to NS following the SN explosion. This fraction of material will begin to rotate into an accretion disk and to fall-back over a long period of time \citep{1989ApJ...346..847C, 2012ApJ...752...32W, 2012MNRAS.419L...1Q}. For simplicity,  we consider a fall-back accretion rate  given by \citep{2018ApJ...857...95M}
\be\label{dotM}
\dot{M} \simeq  \frac23\frac{M_{\rm fb}}{t_{\rm fb}} \cases{ 
1 \hspace{1.6cm}  t \ll t_{\rm fb}  \,, \cr
\left( \frac{t}{t_{\rm fb}} \right)^{-\frac{5}{3}}     \hspace{0.4cm}  t_{\rm fb} \ll t    \,, \cr
}
\ee
where  $M_{\rm fb}$ is the accreting mass  over a characteristic fall-back time $t_{\rm fb}$.   The dynamics of the magnetar with fall-back accretion depends on the Alfv\'en ($r_{\rm m}$), co-rotation ($r_{\rm c}$) and the light cylinder ($r_{\rm lc}$) radii, which are 
\bary\label{rm}
r_{\rm m}&\simeq& 2.2 \times 10^6\,{\rm cm}\, M^{-\frac17}_{\rm ns, 1.4}\,\dot{M}^{-\frac27}_{-2}\,B^{\frac47}_{15}\,R^{\frac{12}{7}}_{\rm ns,6.1}\,,\cr
r_{\rm c}&\simeq& 1.7 \times 10^6\,{\rm cm}\,  M^{\frac13}_{\rm ns, 1.4} \,P^{\frac{2}{3}}_{-3}\,,\cr
r_{\rm lc}&\simeq& 4.8\times 10^6\,{\rm cm}\, P_{-3}\,,
\eary
respectively, and  the spin evolution given by the differential equation  \citep{2011ApJ...736..108P}
\be\label{dif_eq}
I\frac{d\Omega}{dt}=-N_{\rm dip}+N_{\rm acc}\,,
\ee
with $R_{\rm ns}\simeq 1.2\times 10^6\, {\rm cm}\, R_{\rm ns, 6.1}$  the NS radius and $B$ the strength of the dipole magnetic field.   The terms $N_{\rm dip}$ and $N_{\rm acc}$ are the spin-down torques from the dipole emission and accretion, respectively.  For $r_{\rm m}\gtrsim R_{\rm ns}$, these torques are \citep{2016ApJ...822...33P}
\be\label{L_sd}
N_{\rm dip} \simeq   \cases{ 
\frac{\mu^2\Omega^3}{c^3} \frac{r^2_{\rm lc}}{r^2_{\rm m}}  \hspace{0.8cm} r_{\rm m} \lesssim  r_{\rm lc}\,, \cr
\frac{\mu^2\Omega^3}{c^3}   \hspace{1.3cm} r_{\rm lc} \lesssim r_{\rm m}\,, \cr
}
\ee
and
\be\label{Nacc}
N_{\rm acc}=\dot{M}(G\,M_{\rm ns}\,r_{\rm m})^\frac12\, \left[ 1-\left( \frac{r_{\rm m}}{r_{\rm c}} \right)^\frac32 \right]\,,
\ee
where $\mu=BR_{\rm ns}^3$ is the magnetic moment and $G$ is the gravitational constant. In the most general case, the spin-down luminosity can be estimated as $L_{\rm sd}=\Omega(N_{\rm dip}-N_{\rm acc})$.\\  
The magnetar would accrete depending on the location of the Alfv\'en relative to the co-rotation radius. For $r_{\rm m} \lesssim r_{\rm c}$, the magnetar will accrete, otherwise, the system could enter in the propeller regime \cite[e.g., see][]{1998A&ARv...8..279C}. The spin period that delineates both regimes, and also happens to be the steady state evolution of the system  is given for the condition $r_{\rm m}= r_{\rm c}$.  In this case, the spin period in equilibrium becomes 
\be\label{Pc}  
P_{\rm eq}\simeq 1.5\times 10^{-3}\,  {\rm s}\,\, B^{\frac67}_{15}\, R^{\frac{18}{7}}_{\rm ns,6.1}\, M^{-\frac57}_{\rm ns,1.4}\, \dot{M}^{-\frac37}_{-2}\,,
\ee
which is reached during a time interval given by $I\Omega_{\rm eq}/\dot{M}(G\,M_{\rm ns}\,r_{\rm m})^\frac12$ with $\Omega_{\rm eq}=2\pi/P_{\rm eq}$.\\
\subsubsection{GRB prompt emission and the magnetization parameter}
The prompt emission in the Poynting-flux-dominated regime will be generated by the magnetic reconnections which could or not induce internal shell collisions. In both cases, the magnetization parameter plays an important role.  In some magnetic dissipation models,  the magnetization parameter is expected to be similar to the bulk Lorentz factor and lie in the range of $100\lesssim\sigma \lesssim 3000$ \citep{2010ApJ...725.2209L, 2012MNRAS.420..483G, 2017MNRAS.468.3202B}.\\
Irrespective the model, the magnetization parameter is defined by
\be\label{sigma}
\sigma=\frac{L_j}{\dot{M}_j\,c^2}\,,
\ee
where $L_j=L_{\rm sd}$ represents the spin-down luminosity \citep{2009MNRAS.396.2038B} and $\dot{M}_j$ is the rate at which the baryon loading is ejected from the NS surface. In the case of the a weakly-magnetized wind, it can be written as
\be
\dot{M}_j \simeq  \dot{M}_\nu\,f_{\rm cent}\cases{ 
\frac{R_{\rm ns}}{2\,r_{\rm m}}\hspace{0.8cm}   r_{\rm m} \lesssim  r_{\rm lc}\ \cr
\frac{R_{\rm ns}}{2\,r_{\rm m}} \hspace{0.8cm} r_{\rm lc} \lesssim r_{\rm m}\,, \cr
}
\ee
with  $\dot{M}_\nu=\dot{M}_{\rm \nu, ob}(t)+\dot{M}_{\rm \nu, acc}(t)$, $f_{\rm cent}=e^{(\frac{P_{\rm c}}{P})^\frac32}$  and   $P_{\rm c}\simeq 2.7\,R^{\frac12}_{\rm ns}\, r^{-\frac12}_{\rm m}\,M^{-\frac12}_{1.4}$.     The terms $\dot{M}_{\rm \nu, ob}(t)$ and $\dot{M}_{\rm acc}$ are associated with the mass loss rate due to different  sources of neutrinos. The term $\dot{M}_{\rm \nu, ob}(t)$  defined by \citep{2011MNRAS.413.2031M}
\be\label{M_nu-ob}
\dot{M}_{\rm \nu, ob}(t)=3\times10^{-4}\,(1+\frac{t}{t_{\rm kh}})^{-\frac52}\,e^{-\frac{t}{t_{\rm thin}}}\,M_{\odot}\,s^{-1}\,,
\ee
is due to the neutrino ablation,  and 
\be\label{M_nu-acc}
\dot{M}_{\rm acc}(t)=1.2\times10^{-5}\,M_{\rm 1.4}\,\dot{M}_{-2}^{\frac53}\,M_{\odot}\,s^{-1}\,,
\ee
is  due to the accretion \citep{2011ApJ...736..108P}.   The cooling timescale $t_{\rm kh}\approx 2\,{\rm s}$ corresponds to  Kelvin-Helmholtz  and  $t_{\rm thin}\approx (10 - 30)\,{\rm s}$  to the timescale when NS becomes optically thin to neutrinos \citep{2018ApJ...857...95M}.\\
\\
\subsection{Synchrotron light curves from external shocks}
\subsubsection{Light curves  from forward shocks}
It is a widely accepted that the standard synchrotron emission generated by relativistic electrons  accelerated in forward shocks (FSs) can explain the X-ray, optical and radio observations in GRB afterglows. The shape of synchrotron light curves depend on the density profile of the circumburst medium (i.e., uniform-density or wind).  The characteristic and cooling spectral breaks   and the maximum flux of synchrotron emission evolving in a uniform-density (wind) medium are $\epsilon^{\rm syn}_{\rm m}\propto \Gamma^4 (t^{-1} \Gamma^{2})$,  $\epsilon^{\rm syn}_{\rm c}\propto t^{-2}\,\Gamma^{-4}(t\, \Gamma^2)$  and $F^{\rm syn}_{\rm max}\propto t^3 \Gamma^8 (t^0\,\Gamma^2)$, respectively. The predicted synchrotron light curves during the early and late afterglow for the thick and thin shell regime are briefly introduced for an electron distribution of Lorentz factor $\gamma^{-p}_{\rm e}$ with $p$ is the spectral index \citep{1998ApJ...497L..17S}. 
\paragraph{The early afterglow.} During the early afterglow,  before the jet begins to decelerate the bulk Lorentz is constant $\Gamma\propto t^0$ (coasting phase). In this phase, the synchrotron spectral breaks and the maximum flux in the uniform-density (wind) medium are $\epsilon^{\rm syn}_{\rm m}\propto t^0\,(t^{-1})$, $\epsilon^{\rm syn}_{\rm c}\propto t^{-2}\,(t)$ and $F_{\rm max}\propto t^3\,(t^{0})$, respectively. The predicted synchrotron light curves in the uniform-density (wind) medium for the fast- and slow-cooling regime are
{\small
\begin{eqnarray}
\label{fast_early}
F^{\rm syn}_{\rm \nu} \propto \cases{ 
t^2 (t^\frac12) \epsilon_{\rm \gamma}^{-\frac12},\hspace{1.4cm} \epsilon^{\rm syn}_{\rm c}<\epsilon_\gamma<\epsilon^{\rm syn}_{\rm m}, \cr
t^2 (t^\frac{2-p}{2})\epsilon_{\rm \gamma}^{-\frac{p}{2}},\hspace{0.6cm} \hspace{0.45cm} \epsilon^{\rm syn}_{\rm m}<\epsilon_\gamma\,, \cr
}
\end{eqnarray}
}
and
{\small
\begin{eqnarray}
\label{slow_early}
F^{\rm syn}_{\rm \nu} \propto \cases{ 
t^3 (t^\frac{1-p}{2})\, \epsilon_{\rm \gamma}^{-\frac{p-1}{2}},\hspace{1.cm} \epsilon^{\rm syn}_{\rm m}<\epsilon_\gamma<\epsilon^{\rm syn}_{\rm c}, \cr
t^2 (t^\frac{2-p}{2})   \epsilon_{\rm \gamma}^{-\frac{p}{2}},\hspace{0.9cm} \hspace{0.45cm} \epsilon^{\rm syn}_{\rm c}<\epsilon_\gamma\,, \cr
}
\end{eqnarray}
}
respectively.  The synchrotron spectral breaks are reported in \cite{1999A&AS..138..537S, 1999ApJ...520..641S}. \\
\paragraph{The thick-shell regime.} During this regime, this shock is relativistic early.  The bulk Lorentz factor and therefore the synchrotron spectral breaks and the maximum flux in the uniform-density (wind) medium evolves as $\Gamma\propto t^{-\frac14}(t^0)$,  $\epsilon^{\rm syn}_{\rm m}\propto t^{-1}\,(t^{-1})$, $\epsilon^{\rm syn}_{\rm c}\propto t^{-1}\,(t)$ and $F_{\rm max}\propto t\,(t^{0})$, respectively. The predicted synchrotron light curves in the uniform-density (wind) medium for the fast- and slow-cooling regime are 
{\small
\begin{eqnarray}
\label{fast_thick}
F^{\rm syn}_{\rm \nu} \propto \cases{ 
t^{\frac12} (t^{\frac12}) \epsilon_{\rm \gamma}^{-\frac12},\hspace{1.4cm} \epsilon^{\rm syn}_{\rm c}<\epsilon_\gamma<\epsilon^{\rm syn}_{\rm m}, \cr
t^{-\frac{p-2}{2}} (t^{-\frac{p-2}{2}})\epsilon_{\rm \gamma}^{-\frac{p}{2}},\hspace{0.4cm} \epsilon^{\rm syn}_{\rm m}<\epsilon_\gamma\,, \cr
}
\end{eqnarray}
}
and
{\small
\begin{eqnarray}
\label{slow_thick}
F^{\rm syn}_{\rm \nu} \propto \cases{ 
t^{-\frac{p-3}{2}} (t^{-\frac{p-1}{2}})\, \epsilon_{\rm \gamma}^{-\frac{p-1}{2}},\hspace{0.7cm} \epsilon^{\rm syn}_{\rm m}<\epsilon_\gamma<\epsilon^{\rm syn}_{\rm c}, \cr
t^{-\frac{p-2}{2}} (t^{-\frac{p-2}{2}})   \epsilon_{\rm \gamma}^{-\frac{p}{2}},\,\,\,\,\, \hspace{0.75cm} \epsilon^{\rm syn}_{\rm c}<\epsilon_\gamma\,, \cr
}
\end{eqnarray}
}
respectively. The synchrotron spectral breaks are reported in \cite{2013NewAR..57..141G, 2013ApJ...776..120Y}.\\
\paragraph{The deceleration phase (the thin-shell regime).}  In the deceleration phase,  the bulk Lorentz in the uniform-density (wind) medium evolves as $\Gamma\propto t^{-\frac38}\,(t^{-\frac14})$. The synchrotron spectral breaks and the maximum flux in the uniform-density (wind) medium become $\epsilon^{\rm syn}_{\rm m}\propto t^{-3/2}\,(t^{-3/2})$, $\epsilon^{\rm syn}_{\rm c}\propto t^{-\frac12}\,(t^\frac12)$ and $F_{\rm max}\propto t^0\,(t)$, respectively. The predicted synchrotron light curves in the uniform-density (wind) medium for the fast- and slow-cooling regime are 

{\small
\begin{eqnarray}
\label{fast_decel}
F^{\rm syn}_{\rm \nu} \propto \cases{ 
t^{-\frac{1}{4}}\, (t^{-\frac{1}{4}})  \epsilon_{\rm \gamma}^{-\frac12},\hspace{1.4cm} \epsilon^{\rm syn}_{\rm c}<\epsilon_\gamma<\epsilon^{\rm syn}_{\rm m}, \cr
t^{-\frac{3p-2}{4}}\,(t^{-\frac{3p-2}{4}}) \epsilon_{\rm \gamma}^{-\frac{p}{2}},\,\,\,\,\, \hspace{0.35cm} \epsilon^{\rm syn}_{\rm m}<\epsilon_\gamma\,, \cr
}
\end{eqnarray}
}
and
{\small
\begin{eqnarray}
\label{slow_decel}
F^{\rm syn}_{\rm \nu} \propto \cases{ 
t^{-\frac{3p-3}{4}}\,(t^{-\frac{3p-1}{4}}) \, \epsilon_{\rm \gamma}^{-\frac{p-1}{2}},\hspace{0.8cm} \epsilon^{\rm syn}_{\rm m}<\epsilon_\gamma<\epsilon^{\rm syn}_{\rm c}, \cr
t^{-\frac{3p-2}{4}}\,(t^{-\frac{3p-2}{4}}) \epsilon_{\rm \gamma}^{-\frac{p}{2}},\,\,\,\,\, \hspace{0.85cm} \epsilon^{\rm syn}_{\rm c}<\epsilon_\gamma\,, \cr
}
\end{eqnarray}
}
respectively.  The synchrotron spectral breaks are reported in \cite{1998ApJ...497L..17S, 2000ApJ...545..807K, 2000ApJ...536..195C, 2000ApJ...543...66P}.
\subsubsection{Light curves from reverse shocks}
In order to analyze the X-ray/optical flare, the predicted synchrotron light curves in the reverse shock (RS) region are given before and after the shock crossing time 
\paragraph{Before the shock crossing time.}
The synchrotron light curves evolving in the uniform-density (wind) medium for the fast and slow-cooling regime are
{\small
\begin{eqnarray}
\label{fast_before_thin}
F^{\rm syn}_{\rm \nu, r} \propto \cases{ 
 t^{\frac{1}{2}}\,(t^0) \epsilon_{\rm \gamma}^{-\frac12},\hspace{1.4cm} \epsilon^{\rm syn}_{\rm c, r}<\epsilon_\gamma<\epsilon^{\rm syn}_{\rm m, r}, \cr
 t^{\frac{6p - 5}{2}}\,(t^{\frac{p-1}{2}}) \epsilon_{\rm \gamma}^{-\frac{p}{2}},\,\,\,\,\, \hspace{0.4cm} \epsilon^{\rm syn}_{\rm m, r}<\epsilon_\gamma\,, \cr
}
\end{eqnarray}
}
and
{\small
\begin{eqnarray}
\label{slow_before_thin}
F^{\rm syn}_{\rm \nu, r} \propto \cases{ 
 t^{\frac{6p-3}{2}}\,(t^{\frac{p-2}{2}}) \epsilon_{\rm \gamma}^{-\frac{p-1}{2}},\hspace{1.cm} \epsilon^{\rm syn}_{\rm m, r}<\epsilon_\gamma<\epsilon^{\rm syn}_{\rm c, r}, \cr
 t^{\frac{6p - 5}{2}}\,(t^{\frac{p-1}{2}}) \epsilon_{\rm \gamma}^{-\frac{p}{2}},\,\,\,\,\, \hspace{1.05cm} \epsilon^{\rm syn}_{\rm c, r}<\epsilon_\gamma\,, \cr
}
\end{eqnarray}
}
respectively. The sub-index ``r" corresponds to the  spectral breaks and observed flux in the reverse shocks. The synchrotron spectral breaks are reported in \cite{2000ApJ...545..807K, 2000ApJ...536..195C, 2000ApJ...543...66P, 2019ApJ...871..123F}.
\paragraph{After the shock crossing time.}
The synchrotron light curves evolving in the uniform-density (wind) for the fast and slow-cooling regime are
{\small
\begin{eqnarray}
\label{fast_after_thin}
F^{\rm syn}_{\rm \nu, r} \propto \cases{ 
 t^{\frac{-32}{35}}\,(t^{\frac{-11}{14}}) \epsilon_{\rm \gamma}^{-\frac12},\hspace{1.4cm} \epsilon^{\rm syn}_{\rm c, r}<\epsilon_\gamma<\epsilon^{\rm syn}_{\rm m, r}, \cr
 t^{-\frac{27p+5}{35}}\,(t^{-\frac{2 - 13p}{14}})\epsilon_{\rm \gamma}^{-\frac{p}{2}},\,\,\,\,\, \hspace{0.3cm} \epsilon^{\rm syn}_{\rm m, r}<\epsilon_\gamma\,, \cr
}
\end{eqnarray}
}
and
{\small
\begin{eqnarray}
\label{slow_fast_thin}
F^{\rm syn}_{\rm \nu, r} \propto \cases{ 
 t^{-\frac{27p+7}{35}}\,(t^{-\frac{39p+7}{42}}) \epsilon_{\rm \gamma}^{-\frac{p-1}2},\hspace{0.5cm} \epsilon^{\rm syn}_{\rm m, r}<\epsilon_\gamma<\epsilon^{\rm syn}_{\rm c, r}, \cr
 t^{-\frac{27p+5}{35}}\,(t^{-\frac{2-13p}{14}})    \epsilon_{\rm \gamma}^{-\frac{p}{2}},\,\,\, \hspace{0.7cm} \epsilon^{\rm syn}_{\rm c, r}<\epsilon_\gamma\,, \cr
}
\end{eqnarray}
}
respectively. Again, the synchrotron spectral breaks are reported in \cite{2000ApJ...545..807K, 2000ApJ...536..195C, 2000ApJ...543...66P, 2016ApJ...831...22F}.
\subsection{Theoretical Interpretation}
\subsubsection{The initial gamma-ray pulse and the  X-ray precursor}
The  initial gamma-ray pulse is fitted with  an exponential function $\propto \exp(-\frac{t}{\tau_2})$  with $\tau_2=0.39\pm 0.10$ s.  In order to interpret these observations,  we discuss different scenarios such as  an expanding cocoon,  an external shock emission, a high-latitude emission by a relativistic shock breakout and the spin-down magnetar.
\paragraph{Expanding Cocoon.}  \cite{2006ApJ...652..482P} showed that fluxes varying as $t^{-\alpha}$ with $\alpha\approx 2.5 - 4$ could be interpreted as the thermal emission from the expanding cocoon once the jet has broken through the stellar envelope.   In a multi-color BB scenario \citep{2009ApJ...702.1211R}, photons emitted from different angles and with distinct Doppler boosting,  the flux of a thermal spectrum decays as $\propto t^{-2}$.   Given that the thermal emission is not observed in the GBM and the BAT data,\footnote{For BAT analysis see: https://gcn.gsfc.nasa.gov/notices\_s/922968/BA/} we discard the idea that the X-ray precursor and initial gamma-ray pulse could have been produced from an expanding cocoon.\\ 
\paragraph{External shocks.} \cite{2006ApJ...647.1213O} presented the early X-ray observations for 40 bursts using Swift data from BAT and XRT instruments. Authors proposed that the X-ray light curves could be  interpreted as a superposition of the prompt emission and the afterglow. \cite{2018ApJ...859..163H} systematically investigated single-pulse GRBs in the Swift era. Authors found that the prompt emission and the afterglow in a small fraction of bursts originated from external shocks. Some authors have suggested that a single smooth peak or temporally separated peaks during the prompt emission are likely created by external shocks \citep{2004MNRAS.354..915M, 2004ASPC..312..301D, 2016ApJ...822...63B, 2014ApJ...787...90G}. Concerning GRB 190829A,  the initial gamma-ray pulse fits much better with a Band  than the CPL function  and the  BAT observations with a PL than a CPL for a photon index of $\Gamma_{\rm x}=\beta_{\rm x}+1=3.23$. Although  the peak energy evolves as $\propto t^{-\frac32}$, similar to the synchrotron FS model  in the fast cooling regime when the outflow decelerates by  a uniform-density or wind medium, the atypical value of the spectral index $p\approx 5.5$ disagrees with this model\citep[$\beta_{\rm x}=\frac{p-1}{2}$; ][]{1998ApJ...497L..17S}.     Therefore, we discard that  the  X-ray precursor  and initial gamma-ray pulse could have been generated by FSs.  This result is reaffirmed through the analysis of the GRB tail emission (see subsection \ref{tail}) that indicates the prompt emission and the afterglow originate from different components.
\paragraph{High-latitude emission by a relativistic shock breakout.}   \cite{2012ApJ...747...88N} computed the luminosity, the light curve and the spectrum generated by a relativistic breakout, then the planar phase and  finally, the spherical phase. 
They found that spherical relativistic breakouts produce  a gamma-ray flash with an energy, temperature, duration and Lorentz factor well defined and related each other.  In addition, they reported that the predicted flux between the planar and spherical relativistic phase evolves as $\propto t^{-2}$, due to the curvature effect (delayed photons arriving from high latitudes).    Although the evolution of the flux  due to the curvature effect is equal to the initial gamma-ray pulse  a thermal emission is not observed in the GBM and BAT data.  Therefore, we discard that  the  early X-ray observations and the initial gamma-ray pulse can be interpreted by this mechanism.
\paragraph{Spin-down Magnetar.}  The  initial gamma-ray pulse and the X-ray precursor are consistent with the light curve of the spin-down magnetar for a timescale much less than a characteristic fall-back time. Considering the case of  $r_{\rm c}\ll r_{\rm m}$,  the equation (\ref{dif_eq}) becomes
\be
\frac{d\Omega}{dt}+\left( \frac{\mu^2}{c^3 I r^2_{\rm m}}+\frac{\dot{M}\,r^2_{\rm m}}{I} \right) \Omega = 0\,,
\ee
which has as solution $\Omega\propto \, \exp(-\frac{t}{2t_{\rm sd}})$ where the characteristic timescale  is $t_{\rm sd}=\frac12 \left( \frac{\mu^2}{c^3 I r^2_{\rm m}}+\frac{\dot{M}\,r^2_{\rm m}}{I} \right)^{-1}$.  From the evolution of angular frequency, the spin-down luminosity becomes
\bary\label{case_1}
L_{\rm sd}\, &\propto&\,  \exp(-\frac{t}{t_{\rm sd}})\,,
\eary
which has a similar profile to the best-fit curve found for these observations. Considering the typical values $B=10^{16}\,{\rm G}$,  $P=10^{-3}\,{\rm s}$ and $\dot{M}=10^{-2}\,M_{\odot}\, s^{-1}$,  the characteristic timescale is $t_{\rm sd}\approx 2.25\,{\rm s}$. The eq. \ref{case_2} agrees with the best-fit curve of the initial gamma-ray pulse and the  X-ray precursor.
\subsubsection{The ``plateau" phase}\label{plateau}
The X-ray light curve at 1 keV is described by a PL with spectral index $0.06\pm0.05$ that is consistent with a ``plateau" phase.  In order to interpret the ``plateau" phase,  we discuss  different scenarios such as the PL velocity distribution,  the external shock emission and the spin-down magnetar.
\paragraph{A power-law velocity distribution.}  In this scenario the outflow is segmented in several mini-shells each one of them with different velocities.   The equivalent kinetic energy of the outflow is associated with a power-law  velocity distribution given by $E\propto \Gamma^{-\alpha_\Gamma}$ \citep{2001ApJ...551..946T}.     The bulk Lorentz factor in a uniform-density (wind) medium evolves as $\Gamma\propto t^{-\frac{3}{\alpha_\Gamma+8}} (t^{-\frac{1}{\alpha_\Gamma+4}})$ and the predicted flux generated by  synchrotron emission evolves as  $F_\nu\propto t^{\frac{3(\alpha_\Gamma-2p+2)}{\alpha_\Gamma+8}}$ ($t^{\frac{\alpha_\Gamma-6p+2-\alpha_\Gamma p}{2(\alpha_\Gamma+4)}}$) for $\epsilon^{\rm syn}_{\rm m}<\epsilon_\gamma<\epsilon^{\rm syn}_{\rm c}$ and  $F_\nu\propto t^{\frac{2(\alpha_\Gamma-3p+2)}{\alpha_\Gamma+8}}$ ($t^{\frac{2\alpha_\Gamma-6p+4-\alpha_\Gamma p}{2(\alpha_\Gamma+4)}}$) for $\epsilon^{\rm syn}_{\rm c} < \epsilon_\gamma$ \citep[e.g, see][]{2000ApJ...535L..33S, 2006ApJ...642..354Z, 2015MNRAS.448..417B, 2019ApJ...871..200F}. In order to reproduce the ``plateau" phase, an atypical value of $p\approx4.0 (1.0)$ has to be required for the uniform-density (wind) medium. This is very far from expectations from particle acceleration.
\paragraph{Synchrotron FS emission.}   In this scenario, the ``plateau" phase is interpreted as synchrotron FS emission during the thick-shell regime (before the significant  deceleration begins). From eqs. \ref{fast_thick} and \ref{slow_thick} can be seen that the synchrotron FS model evolving in a uniform-density medium  can describe these observations for $p=3.2\pm0.2$ ($F_\nu\propto t^{0.1\pm0.1}$; $\epsilon^{\rm syn}_{\rm m}<\epsilon_\gamma<\epsilon^{\rm syn}_{\rm c}$), $p=2.1\pm0.2$ ($F_\nu\propto t^{0.05\pm0.10}$; $\epsilon^{\rm syn}_{\rm c} < \epsilon_\gamma$  ) and in the wind medium for  $p=1.3\pm0.2$ ($F_\nu\propto t^{0.1\pm0.1}$; $\epsilon^{\rm syn}_{\rm m}<\epsilon_\gamma<\epsilon^{\rm syn}_{\rm c}$).  Due to the atypical value required for the wind medium, again we discard it.\\ 
For an spectral index of $p=3.2\pm0.2$,  the X-ray and optical light curves would evolve in the same PL segment and for $p=2.1\pm0.2$ these bands would evolve in distinct PL segments.   Because the X-ray and optical light curves are normalized at $\epsilon_\gamma$=1 keV and $\sim $ (1 - 3) eV, respectively,  the predicted optical flux must be much larger than the X-ray flux  ($F_{\rm \nu, opt}/F_{\rm \nu, X}\gg1$), for typical values of parameters.   However, during this phase the observed fluxes in the White and U bands are less than the observed X-ray flux. Therefore, we discard the synchrotron FS emission as origin of the ``plateau" phase.\\
\paragraph{Spin-down Magnetar}   Once the system reaches the equilibrium  ($r_{\rm c}= r_{\rm m}$),  the accretion term is zero  ($N_{\rm acc}(\Omega_{\rm eq})=0$), and therefore the spin-down luminosity becomes  $L_{\rm sd}=\Omega N_{\rm dip}$.  From eqs. (\ref{L_sd}) and (\ref{dotM}),  the spin-down luminosity is in the form
\bary\label{case_2}
L_{\rm sd}&\simeq&\frac{\mu^2\,\Omega^4_{\rm eq}}{c^3}\frac{r^2_{\rm lc}(\Omega_{\rm eq}) }{r^2_{\rm c}}\propto \cases{ 
 t^{0}\hspace{1.cm}  t \ll t_{\rm fb}  \,, \cr
t^{-\frac{50}{21}} \hspace{0.6cm}  t_{\rm fb} \ll t    \,, \cr
}
\eary
which  for  $t \ll t_{\rm fb}$ is consistent with the best-fit curve during this phase.  It is worth noting that the best-fit value of $0.573\pm0.013$ reported by the Swift analysis\footnote{https://www.swift.ac.uk/xrt\_live\_cat/00922968/1\_breaks.png} during the time interval  $\sim (10^2 - 10^5)\,{\rm s}$ would agree with the light curve given by eq. (\ref{case_2}) during the transition around $t_{\rm fb}$. \\ 
\\
Other mechanisms such as a photospheric component from a moderate outflow injected after the prompt emission \citep{2017A&A...605A..60B} and viewing of bursts very slightly off-axis \citep{2020MNRAS.492.2847B} have been recently proposed to interpret the  ``plateau" phase.  However,  a photospheric component was not clearly observed in this burst and the $E_{\rm peak}$ and $E_{\rm iso}$ relation indicates that this burst was on-axis (see subsection \ref{Epeak-Eiso}), so these mechanism are discarded.\\ 
\subsubsection{A normal decay phase}
After the break time of $1.4^{+0.17}_{-0.15}  \times 10^5\,{\rm s}$, the best fit of the X-ray observations reported was  $\alpha_{\rm X,IV}=1.24\pm0.04$.\footnote{https://www.swift.ac.uk/xrt\_live\_cat/00922968/} Before the break time, a temporal index of $\alpha_{\rm X,IV}=1.05\pm0.02$ was obtained.   These observations are consistent with the synchrotron FS model evolving in the slow-cooling regime and in a uniform-density medium for $p\approx 2.3$. During  this phase,  the hardness ratio light curve shows a very moderate spectral softening.\\   
Spectral analysis at $1.8\times 10^4\,{\rm s}$ indicates that the X-ray and optical observations evolve in the same PL segment  with spectral slope of $\beta_{\rm X\cdot O}=0.47\pm0.09$. This value is consistent with the PL segment of the  slow-cooling regime  $\epsilon^{\rm syn}_{\rm m}<\epsilon_\gamma<\epsilon^{\rm syn}_{\rm c}$  for $p=1.98\pm0.05$. Otherwise, an atypical value of $p$ is obtained where X-ray and optical observations would evolve in distinct PL segments.\\
Taking into account the temporal and spectral analysis (the closure relations given by eq. \ref{slow_decel}), this phase is consistent with synchrotron FS model that evolves during the deceleration phase in a uniform-density medium for $p=2.15\pm0.17$.  The value of $p$ is consistent with the range of values typically for GRB afterglows  \citep{2015PhR...561....1K}.\\
\subsubsection{The X-ray and optical flares}\label{flares}
The X-ray and optical flares occurred simultaneously peaking at $\sim 1.4\times 10^3\,{\rm s}$ with rising and falling slopes of $\approx -3$ and $\approx 1$, respectively.    In what follows we discuss the late central-engine activity, the neutron signature and the synchrotron emission from external shocks as possible scenarios.
\paragraph{Late central-engine activity.}
In the fireball model,  faster shells in an emitting region of the jet interact with slower ones.  If the comoving magnetic field in the emitting region is random or transverse, the flux would evolve as $t^{-\alpha}$ with $\alpha=2(p+1)$ or $\frac{2-3p}{2}$, respectively, if the synchrotron emission evolves in the fast-cooling regime for $\beta=\frac{p}{2}$, and  $\alpha=2p$ or $\frac{1-3p}{2}$, respectively, in the slow cooling regime for $\beta=\frac{1-p}{2}$.   Given the best-fit values of the falling slopes of the X-ray and optical flares (see Tables \ref{table3:X_ray} and \ref{table4:optical}),  an atypical value of $p<1$ would have to be required to reproduce these flares.    As the falling slopes are less than $<2$, these cannot be interpreted in the context of high-latitude emission. Considering  the simultaneity of the X-ray and optical flares and that the best-fit values of the timescale lie in the range of 0.68 - 0.9 and,  internal shocks created by late central-engine activity cannot reproduce the features of these flares. It is important to highlight that  the photon index and the hardness ratio displayed random fluctuations around $\approx 3$ and  $\approx 2.5$, respectively, different to the behavior during the flaring activity (see section \ref{evolution}).\\
\paragraph{Neutron Signature.}
A relevant process that could describe the X-ray and optical flares is associated with the presence of neutrons in the outflow \citep{1999ApJ...521..640D, 2014ApJ...787..140F}. It occurs as neutrons and ions are fully decoupled; neutrons create a leading front and ions start to slow down. Then, neutrons decay in products that interact with the slow-moving ions and produce a re-brightening in the early afterglow \citep{2003ApJ...585L..19B}.  \cite{2005MNRAS.364L..42F}
introduced an analytic formalism to derive the light curves and spectrum.   Authors showed that the resulting light-curve initially increases fast, then there comes a flat phase the duration of the burst and finally, it drops sharply.  Although, the light curve by this scenario has a similar profile  to those displayed in GRB 190829A, this occurs at early times ($\sim T_{90}$).  Given that the  X-ray and optical flares peak at $\sim 1.4\times 10^3\,{\rm s}$, we conclude that neutron-proton decoupling cannot describe these flares exhibited in GRB 190829A.
\paragraph{Synchrotron RS emission.} 
Synchrotron emission from RSs are usually required to describe X-ray and optical flares \citep{2000ApJ...545..807K, 2016ApJ...818..190F, 2017ApJ...848...15F, 2018ApJ...859...70F}.   From eqs.  (\ref{fast_before_thin}) -  (\ref{slow_fast_thin})  can be seen that before and after the shock crossing time, the synchrotron RS model evolving in slow-cooling regime and uniform-density medium  could describe the temporal and spectral observations but for an atypical value of $p\approx 1.3\pm0.2$.     Similarly, the fact that the soft X-ray and optical fluxes are simultaneous during more than $\sim 10^4\,{\rm s}$ with similar slopes disfavor the synchrotron  RS scenario. It is worth noting that the predicted synchrotron light curves evolving in a wind medium cannot explain  the rising slopes.\\  
\paragraph{Synchrotron FS emission.}
At early times,  the optical and X-ray light curves show bright simultaneous flares peaking at $\sim 1.4\times 10^3$ s with variability timescales in the range of $\Delta t/t \approx 0.68 - 0.9$.   Tables \ref{table4:optical} and \ref{table3:X_ray} show that  during the time interval $\sim (0.65 - 1.4)\times 10^3$ s,  the X-ray and optical fluxes increase as $t^{-\alpha}$  with slopes of $\approx -3$,  reaching a maximum flux  at $\sim1.4\times 10^3\,{\rm s}$.  The closure relations during the early afterglow/coasting phase (eqs. \ref{fast_early} and  \ref{slow_early}) indicate that the synchrotron FS model is the slow-cooling regime ($\epsilon^{\rm syn}_{\rm m}<\epsilon_\gamma<\epsilon^{\rm syn}_{\rm c}$) and evolving  in a uniform-density medium  could describe these observations.   It is worth noting that the predicted synchrotron light curves evolving in a wind medium cannot reproduce the rising slopes for any value of the electron spectral index. In the deceleration phase, the predicted synchrotron light curves decrease as $t^{-\alpha}$ with $\alpha \approx 1$ for p=2.3 which is consistent once the X-ray and optical fluxes begin to decrease (see eqs. \ref{fast_decel} and  \ref{slow_decel}).  
On the other hand, spectral analysis at $6\times 10^3\,{\rm s}$ and $1.8\times 10^4\,{\rm s}$ indicate that the X-ray and optical observations evolve in the same PL segment  with a spectral slope of $\beta_{X\cdot O}=0.48\pm0.05$ and $0.47\pm0.09$, respectively. These values are consistent with the PL segment of slow-cooling regime  $\epsilon^{\rm syn}_{\rm m}<\epsilon_\gamma<\epsilon^{\rm syn}_{\rm c}$  for $p\approx 2$. \\
Finally,  the temporal and spectral analysis indicates that the X-ray and optical flares can be theoretically described as synchrotron FS emission for $ p\approx 2.15\pm 0.17$.\\
\\
It is worth noting that photospheric emission from moderate Lorentz factor material that is emitted at the same time as the material producing the prompt leads to much weaker flares in the optical compared to X-rays, so it may not be appropriate here \citep{2016MNRAS.457L.108B}.\\ 
\subsection{Modelling the multi-wavelength light curves}
\begin{figure}[h!]
{\centering
\resizebox*{0.9\textwidth}{0.4\textheight}
{\includegraphics{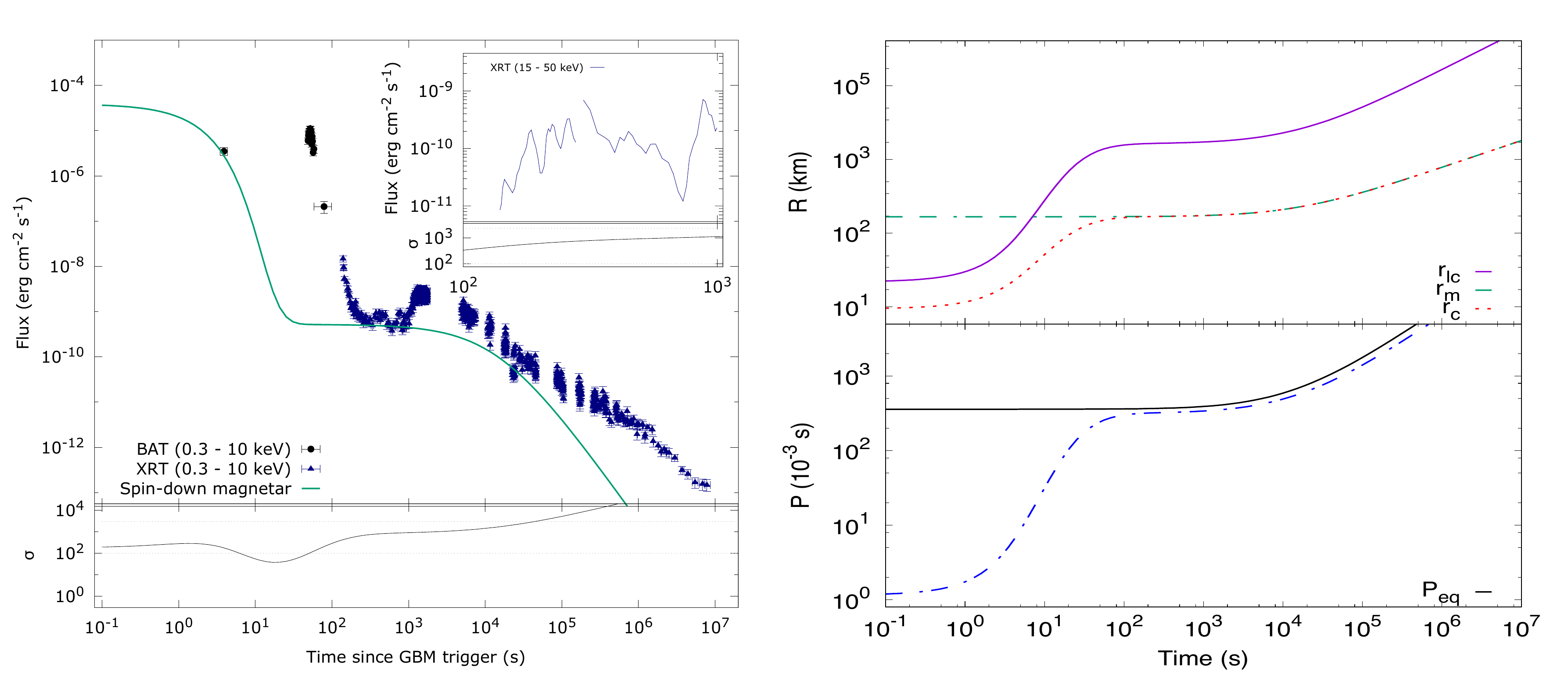}}
\resizebox*{0.9\textwidth}{0.5\textheight}
{\includegraphics{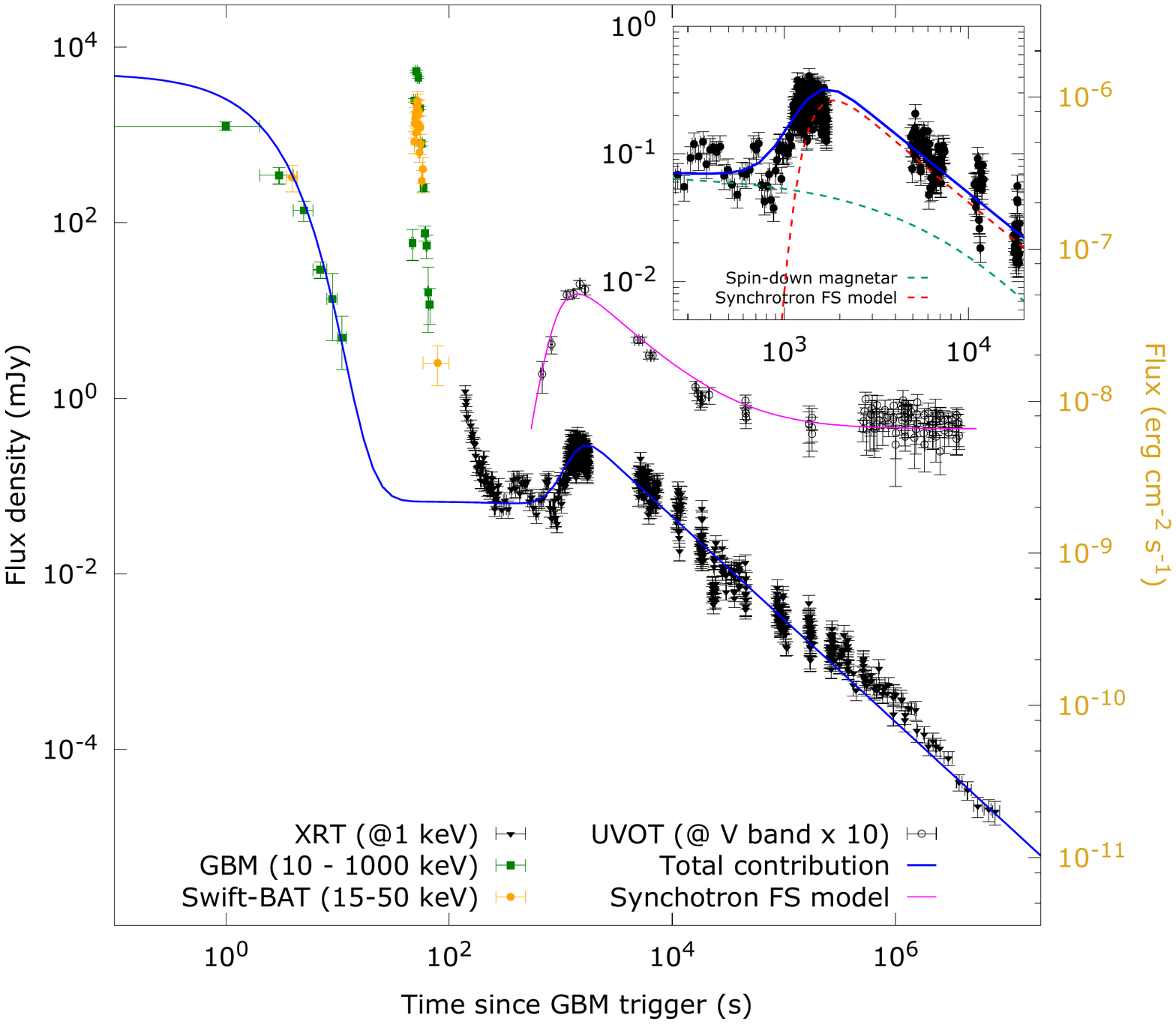}}
}
\caption{Upper left-hand panel shows the X-ray light curve with the best-fit curve given by the spin-down magnetar (above) and the evolution of the magnetization parameter (above). The small box shows the lightcuve at 10 keV.  The upper right-hand panel shows the evolution of the Alfv\'en ($r_{\rm m}$), co-rotation ($r_{\rm c}$) and the light cylinder ($r_{\rm lc}$) radii (above) and the evolution of the spin period (below). The lower panel shows the multi-wavelength light curves of GRB 190829A with the best-fit curves given by the spin-down magnetar and the synchrotron FS model.}
\label{figure3}
\end{figure}

The spin-down luminosity can be converted  into isotropic X-ray flux through the efficiency in converting its spin-down energy to radiation ($\eta_{\rm x}$)  and the beaming factor of the magnetar wind ($f_b=1-\cos\theta_j$). The  X-ray luminosity can be written as 
\be\label{Lx}
L_{\rm x}=\eta_{\rm x}\,f_b^{-1}\,L_{\rm sd}\,.
\ee
Similarly, considering the typical energy range of Swift XRT and the photon spectral index  for GRB 190829A,   the X-ray luminosity could be converted to X-ray flux at 1 keV using the relationship $F_{\rm x}\simeq \frac{L_{\rm x}}{4\pi d^2_{\rm z}}$.  The term $d_{\rm z}$ corresponds to the luminosity distance which is estimated using the values of the Hubble constant as $H_0=(67.4\pm 0.5)\,{\rm km\,s^{-1}\,Mpc^{-1}}$ and the matter density parameter as $\Omega_{\rm m}=0.315\pm 0.007$ \citep{2018arXiv180706209P}. \\
The upper panels in Figure \ref{figure3} show the X-ray (BAT and XRT) light curve at (0.3 -10) keV energy range with the best-fit curve given by the spin-down magnetar (left) and the evolution of the Alfv\'en ($r_{\rm m}$), co-rotation ($r_{\rm c}$) and the light cylinder ($r_{\rm lc}$) radii (right).   Swift data in the small box included in the upper left-hand panel is reported at 10 keV.    The fit was done using the MINUIT algorithm \citep{James:1975dr} via the \texttt{iminuit}\footnote{A python interface to minuit. Accessed: 2018-03-05.  \url{https://github.com/scikit-hep/iminuit}} Python interface. We use the interpolation of the solution using the \texttt{interp1d} function from the \texttt{scipy.interpolate} Python object and the $\chi^{2}$ regression function with the \texttt{Chi2Regression} object from the \texttt{probfit} Python interface.  A set of initial values  are placed in order to minimize the function with the \texttt{iminuit} interface and the \texttt{migrad} optimiser.   We assume a fall-back mass of $M_{\rm fb}=0.8 M_\odot$ \citep{2018ApJ...857...95M}, $\theta_j=8^\circ$ (see analysis in section \ref{tail}) and $\eta_{\rm x}=0.1$ \citep{2013ApJ...775...67B}.   The best-fit values of the magnetic field, the spin period and the fall-back timescale are reported in Table \ref{table:parameters}.   This panel shows the evolution of the magnetization parameter (below).    This parameter lies at two separate time intervals in the range of $ 100\lesssim \sigma\lesssim 3000$;  before a few seconds and larger than $\sim 80\,{\rm s}$. The small box shows that evolution of the magnetization parameter agrees with the flaring activity at (15 -50) keV energy range.    During the early X-ray observations the extracted rotational energy is $4.18\times 10^{50}\,{\rm erg}$ and during the ``plateau" phase is $2.93\times 10^{47}\,{\rm erg}$.  Comparing the extracted rotational energy associated to the precursor with the isotropic-equivalent energy (see Table \ref{table1:gbm_analysis}),  we can estimate an  efficiency of ($21.7\pm1.13$)\%.  The upper right-hand panel shows  that the spin period reaches its equilibrium at $\sim 10\,{\rm s}$ (on a timescale of$\sim10^{-3}\,t_{\rm fb}$). During the first seconds, the light cylinder radius  is less than Alfv\'en radius,  so the spin-down luminosity is similar to the isolated magnetar.  The system lies in the propeller regime so that the angular momentum losses decrease the total rotational energy.\\
\\
The lower panel shows the multi-wavelength light curves of GRB 190829A with the best-fit curves given by the spin-down magnetar and the synchrotron FS model.  The UVOT data  is exhibited at the V band.      The quantities observed for GRB 190829A such as the redshift $z=0.078$, the isotropic-equivalent energy  $2.967\times 10^{50}\,{\rm erg}$ and   the spectral index $\Gamma_{\rm x}=2.2$\footnote{https://www.swift.ac.uk/xrt\_spectra/00922968/} are used. The solid line in blue shows the total contribution (the spin-down magnetar and the synchrotron FS model, see the small box) and in magenta show the synchrotron FS model with the host galaxy contribution. The best-fit values of the uniform-density medium ($n$), the equivalent-kinetic energy ($E$), the spectral index ($p$) and the microphysical parameters given to accelerate electrons ($\varepsilon_{\rm e}$) and  to amplify the magnetic field ($\varepsilon_{\rm B}$) during the FSs are reported in Table \ref{table:parameters}.  Given the parameters found with our model we can estimate: i) the value of the initial bulk Lorentz factor to be $\Gamma\simeq 34$ which is low for a  hlGRB and high for a  llGRB \citep{2014A&A...566A.102S},  ii) the deceleration radius at the peak time ($1.4\times 10^3\,{\rm s}$) of the afterglow is $R_{\rm dec}\simeq 0.05\,{\rm pc}$, iii) the efficiency to convert  the kinetic energy into photons is $\eta_{\rm k}=12.4$ \% which is typical for GRB afterglows  \citep{2015PhR...561....1K,2015MNRAS.454.1073B} and iv)   the spectral break energies are $\epsilon^{\rm syn}_{\rm m}=1.9\,{\rm eV}$ and $\epsilon^{\rm syn}_{\rm c}=0.1\,{\rm MeV}$ which indicate that synchrotron model evolves in the slow-cooling regime. \\
Based on the analysis and modelling of the multi-wavelength observations, we present a full discussion of GRB 190829A 
\vspace{0.4cm}
\section{Results and Discussion}
\subsection{Non-existence of the thermal emission}
Evoking the photosphere, the shock breakout and the cocoon models,  the thermal emission is expected during the precursor and the main emission episode.  For GRB 190829A,  the  Band function  fits better the GBM  data (the initial pulse and the brighter peak) than the BB function, and  the PL function fits the BAT data better than the BB function.\footnote{For BAT analysis see: https://gcn.gsfc.nasa.gov/notices\_s/922968/BA/}   The non-detection of a clear thermal emission in both the GBM and BAT data indicates that GRB jet outflow might be dotted with a significant fraction of magnetic field \citep{2009ApJ...700L..65Z, 2015ApJ...801..103G}  or the Band  function is  the result  of the  reprocessed  quasi-thermal emission from kinetic or magnetic dissipation processes near the photosphere \citep{2006ApJ...652..482P, 2010ApJ...725.1137L, 2016ApJ...831..175V, 2015MNRAS.454L..31A, 2013MNRAS.428.2430L, 2011ApJ...738...77V, 2012ApJ...761L..18V}. In the standard fireball model, the prompt emission is expected to be a superposition of non-thermal and quasi-thermal photosphere components as found in the BATSE  \citep{2009ApJ...702.1211R}  and Fermi \citep{2010ApJ...709L.172R, 2015ApJ...814...10G} data.  The detection of a single non-thermal emission as well as the evolution of the spectral parameters (see subsection \ref{evol})  would favor the scenario of a Poynting-flux-dominated outflow \citep{2009ApJ...700L..65Z, 2015ApJ...801..103G,  2014MNRAS.445.3892B}. \\
\subsection{The brighter gamma-ray peak}\label{brighter}
Analysis of the brighter peak indicates that the gamma-ray flux decays as $F_\nu\propto t^{-6.58\pm1.26}$. This decay can be interpreted in terms of internal shocks.  During  these shocks, a large fraction of Poynting flux is converted into kinetic energy and dissipated at larger radii.   Taking into consideration the temporal decay index of $6.58\pm1.26$,  a random magnetic field in the emitting region for $p= 3.29\pm0.63$ is favored.  Similarly, the peak energy evolves as a function of time $\propto t^{-\alpha_{\rm peak}}$ with a temporal index of $\alpha_{\rm peak}\sim 1$ which is consistent with internal shocks  \citep[e. g., see][]{2011MNRAS.413.2031M, 2017MNRAS.472.3058B}.   It is worth noting that the internal shock scenario agrees with the low- and high-energy spectral index $\alpha\approx -1.1$ and $\beta\approx -2.5$, respectively.   The value of low-energy spectral index of $\sim - 1$ is consistent with  the synchrotron radiation in magnetic dissipation models that require large dissipation radius \citep[$\sim 10^{15}\,{\rm cm}$;][]{2014NatPh..10..351U, 2016MNRAS.459.3635B,2018MNRAS.476.1785B}.\\  
\\
\subsection{Analysis of the GRB tail emission}\label{tail}
The GRB tail emission marks the end of the prompt phase and also the onset of the afterglow \citep{2006ApJ...642..354Z}. This emission reveals whether the prompt emission and the afterglow originate from distinct components or emitting sites.    If the prompt emission and the afterglow arise from distinct components or emission sites,  an abrupt decay in the flux level should be observed during the transition phase  between  the prompt emission and  the afterglow. Such an abrupt decay which accounts for the delayed photons is associated with the high-latitude emission due to the curvature effect. \cite{2000ApJ...541L..51K} showed that the observed flux associated to high latitudes evolves as $F_{\rm \nu}\propto t^{-(2+\beta)}$, where $\beta$ refers to the spectral index of the synchrotron emission, $\beta=\frac12$ or $\frac{p}{2}$ for fast- and $\beta=\frac{1-p}{2}$ or $\frac{p}{2}$ for slow-cooling regime.  It is worth  mentioning that some variations to the curvature effect model including a non-power-law spectrum \citep{2009ApJ...690L..10Z}, an anisotropic emission \citep{2011MNRAS.410.2422B, 2016MNRAS.455L...6B,2016MNRAS.459.3635B}, a structured jet \citep{2014ApJ...787...90G}, etc., that leads to a different flux evolution during this phase have been proposed.\\
 During the interval 82 $ \lesssim t \lesssim 215\,{\rm s}$,  the X-ray light curve at 1 keV exhibited a fast decay which is difficult to explain by invoking the external shock model.  Some authors have argued that central engines cannot die suddenly, and that the observed GRB tails may give account about the dying history of the central engines \citep{2005MNRAS.364L..42F, 2009MNRAS.395..955B}.   Given the best-fit value of the steep decay of the GRB tail emission $\alpha_{X, I}=3.53\pm0.70$ and the strong variation exhibited by the spectral index and the hardness ratio, the X-ray observations in GRB 190829A during this epoch can be interpreted in the context of the high-latitude emission for $p= 3.06\pm 1.40$ \citep{2000ApJ...541L..51K}.  This value is consistent with the spectral index ($p= 3.29\pm0.63$) derived from the brighter gamma-ray peak (see subsection \ref{brighter}).   The hardness ratio light curve shows a soft-to-hard spectral evolution. As seen in other bursts detected by Swift \citep{2007A&A...468..103G},   a temporal analysis  shows that the photon index and source intensity are highly correlated with the spectrum being harder when brighter.   This result clearly indicates  that the steep decay observed in the Swift XRT data is related to the prompt emission \citep{2005ApJ...635L.133B, 2005Natur.436..985T}.\\
Given the GRB tail emission, the opening angle for a canonical jet can be estimated as
\be
\theta_{j}\approx 8^\circ\,\left(\frac{1+z}{1.08}\right)^{-\frac12}\,t_{\rm tail, 2}^{ \frac12}\,R_{\rm cr,14.5}^{-\frac12}\,, 
\ee
where  $t_{\rm tail}$ is the duration of the tail and $R_{\rm cr}$ is the curvature radius.  Using the opening angle,  the initial bulk Lorentz factor is $\Gamma\gtrsim 25 \frac{t^\frac12_{\rm tail, 2}}{\sqrt{1-\cos\theta_j}}$ which  is consistent with the value estimated from peak time of the afterglow.\\ 
Before and after the GRB tail emission we identity the prompt emission and afterglow, respectively.  A rough comparison between the X-ray emission level during the prompt emission and the afterglow  could indicate if both emissions originated in the same component  \citep{2006ApJ...642..354Z}. Given the low-energy spectral index $\alpha_{\rm Band}$, the energy peak  and the total isotropic energies reported in Table 1, the duration of the burst, the equivalent kinetic energy ($E_{\rm k}=10^{52}\,{\rm erg}$), the energy at which the X-rays are reported ($E_{\rm X}=1 \,{\rm keV}$)  and microphysical parameter ($\varepsilon_e\approx 0.1$),  the flux ratio is  \citep[see,][]{2006ApJ...642..354Z}
\be
\frac{F^{\rm pr}_{\nu, X}}{F^{\rm ag}_{\nu, X}}\approx 0.4 \left(\frac{E_{\rm \gamma, iso, 51}}{E_{\rm k, 52}}\right) \left(\frac{t_2}{T_{90}}\right) \left( \frac{E_{\rm X, 1\,keV}}{E_{\rm pk, 10\, keV}} \right)^{\alpha_{\rm Band}+2}\varepsilon^{-1}_{\rm e,-1}\,,
\ee
which is in stark contrast with the flux ratio ($>10^4$) observed in GRB 190829A.     This result agree with our model where  the prompt emission and the afterglow originate from different components.\\ 
\subsection{$E_{\rm peak}$ - $E_{\rm \gamma, iso}$ Relation}\label{Epeak-Eiso}
\begin{figure}[h!]
{\centering
\resizebox*{0.8\textwidth}{0.5\textheight}
{\includegraphics{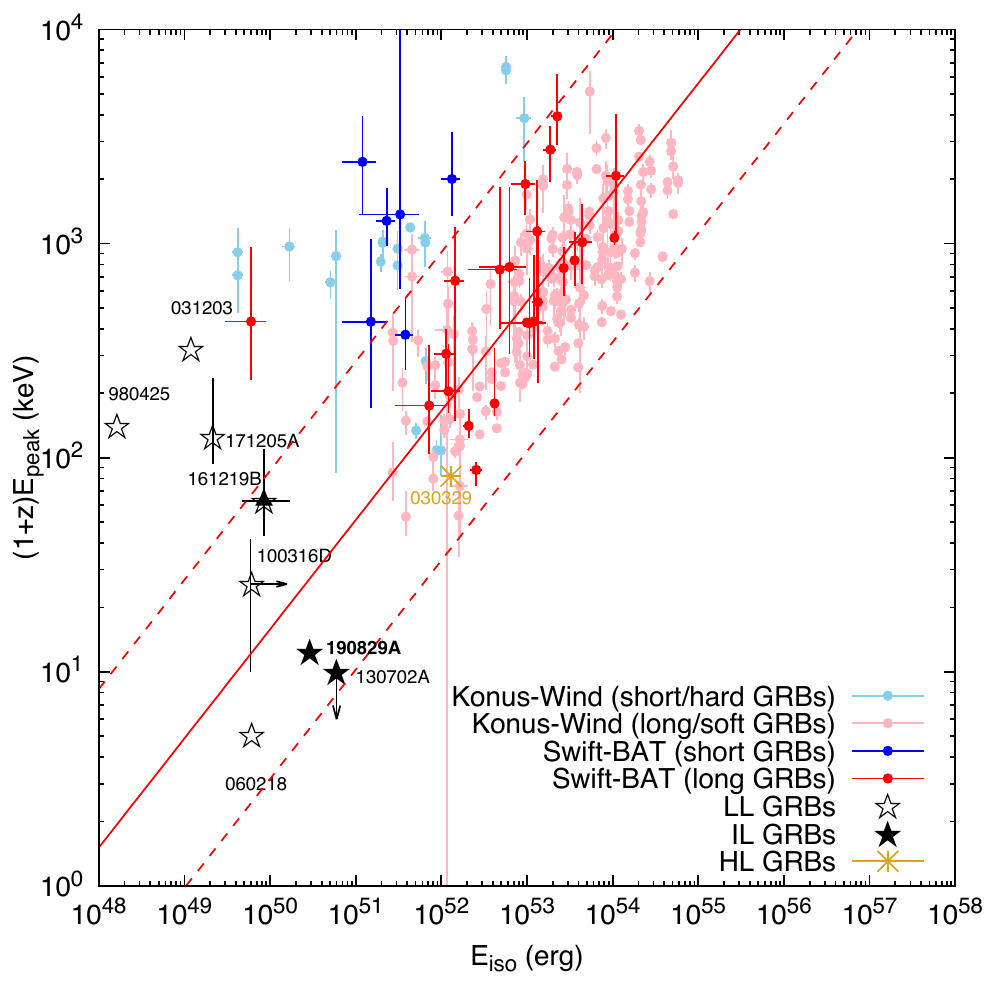}}
}
\caption{The $E_{\rm peak}$ and $E_{\rm iso}$ relation for GRB detected by Konus-Wind, Swift BAT and low-luminosity GRB samples. This figure is adapted from \cite{2018A&A...619A..66D}. The black filled star in bold face shows GRB 190829A. The solid and dashed red lines correspond to the best fit and  vertical  logarithmic  deviations ($2.5\sigma$), respectively, reported in \citep{2009ApJ...704.1405K}.  The $E_{\rm peak}$ and $E_{\rm iso}$ of LL, IL and HLGRBs are obtained from  GRB 980425/ SN 1998bw \citep{1998Natur.395..670G},  GRB 060218/ SN 2006aj \citep{2006Natur.442.1008C},  GRB 100316D/ SN 2010bh \citep{2011ApJ...740...41C},  GRB 161219B/SN 2016jca \citep{2017A&A...605A.107C, 2019MNRAS.487.5824A}, GRB 171205A/ SN 2017iuk \citep{2019Natur.565..324I},   GRB 130702A/ SN 2013dx \citep{2015A&A...577A.116D} and  GRB 030329/SN 2003dh  \citep{2003Natur.423..847H}.}
\label{figure4}
\end{figure}

Using the derived values of the total isotropic energy $(2.967\pm 0.032)\times 10^{50}\,{\rm erg}$ and the peak energy of the gamma-ray spectrum $E_{\rm peak}=11.47 \pm 0.360\, {\rm keV}$, we found that GRB 190829A is consistent with the $E_{\rm peak}$ - $E_{\rm \gamma, iso}$ relation \citep[Amati relation;][]{2002A&A...390...81A, 2006MNRAS.372..233A} as other GRB-SNe associated with  millisecond magnetars \citep{2014A&A...568A..19C}. Figure \ref{figure4} shows the $E_{\rm peak}$ and $E_{\rm \gamma, iso}$ relation for GRBs detected by Konus-Wind, Swift BAT and the low-, intermedium-  and high-lGRB sample within $z\lesssim 0.2$.   The best-fit value of the correlation is adapted from \cite{2018A&A...619A..66D}. The low-, intermedium-  and high-lGRBs are obtained from  GRB 980425/ SN 1998bw \citep{1998Natur.395..670G},  GRB 060218/ SN 2006aj \citep{2006Natur.442.1008C},  GRB 100316D/ SN 2010bh \citep{2011ApJ...740...41C},  GRB 161219B/SN 2016jca \citep{2017A&A...605A.107C, 2019MNRAS.487.5824A}, GRB 171205A/ SN 2017iuk \citep{2019Natur.565..324I},   GRB 130702A/ SN 2013dx \citep{2015A&A...577A.116D} and  GRB 030329/SN 2003dh  \citep{2003Natur.423..847H}.   It worth noting that GRB 190829A follows this relation, indicating that this burst was not off-axis.   
%
%
\vspace{1cm}
\section{Analysis and discussion of the multi-GeV photons} 
\subsection{Synchrotron limit}
\begin{figure}[h!]
{\centering
\resizebox*{0.48\textwidth}{0.33\textheight}
{\includegraphics{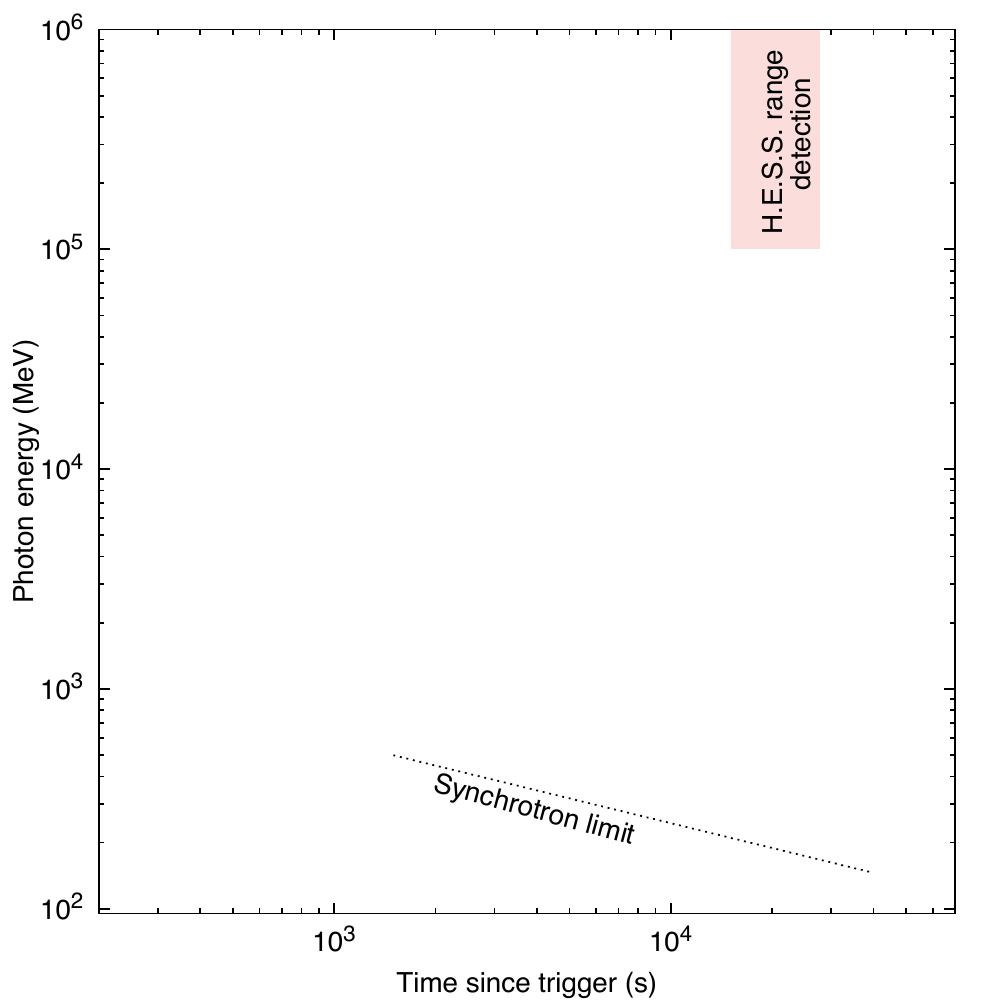}}
\resizebox*{0.48\textwidth}{0.33\textheight}
{\includegraphics{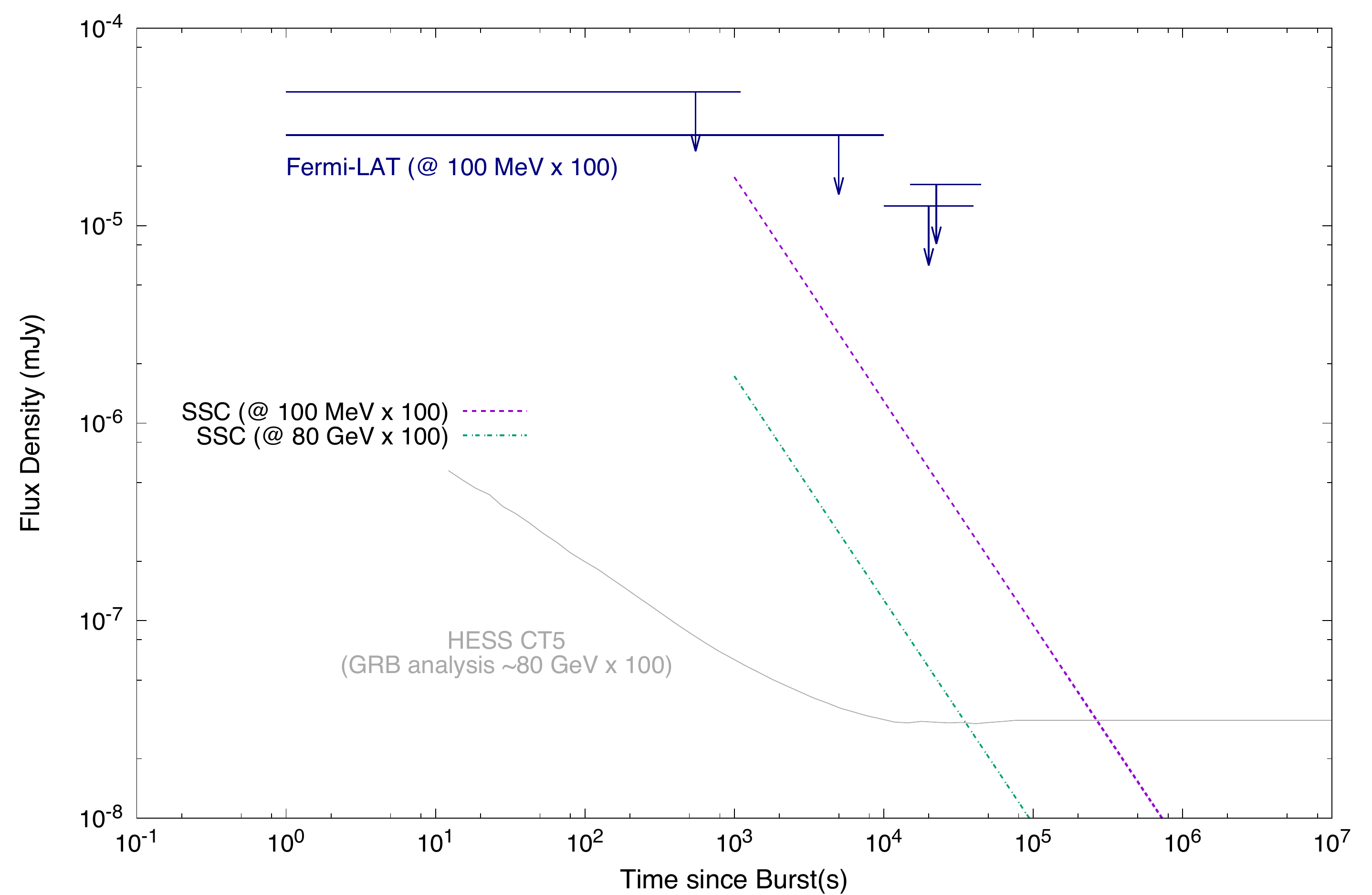}}\\
}
\caption{The left-hand panel shows the interval and the energy range of VHE photons reported by the H.E.S.S. Collaboration (the pink region) and the synchrotron limit (the dashed line).   The right-hand panel  shows the SSC emission estimated at 100 MeV and 80 GeV from the forward- and reverse-shock model computed from the deceleration of the wide jet in a circumburst medium with uniform density. In order to verify our model with the observations at high and very-high energies,  the Fermi-LAT upper limits at 100 MeV and the sensitivity of H.E.S.S. at 80 GeV. \citep{2016CRPhy..17..617P}. }
\label{figure5}
\end{figure}

Given the best-fit parameters (see Table \ref{table:parameters}), we plot in the left-hand panel in Figure \ref{figure5} the evolution of the maximum energy photon radiated by the synchrotron FS model  in uniform-density medium.   In addition,   we plot  the VHE emission in the range reported by H.E.S.S..  It is shown that the VHE photons  cannot be interpreted in the synchrotron FS scenario.  Therefore,  an additional mechanism should be present during the FSs to explain these multi-GeV photons.  However, photons at hundreds of MeVs  below the synchrotron limit can be explained in terms of synchrotron FS scenario \citep{2009MNRAS.400L..75K}.\\
Synchrotron photons radiated in the FSs can be up-scattered by the same electron population.  Here, we use the best-fit parameters reported in Table \ref{table:parameters} and  the  SSC emission during the deceleration phase in a uniform-density medium introduced in \cite{2019ApJ...883..162F}.
\subsection{Synchrotron-self Compton FS model}
Once the jet begins decelerating, the intrinsic attenuation due to $\gamma\gamma$ interactions\footnote{VHE gamma-ray photons can interact with lower-energy photons to produce pairs} can be estimated by \cite[e.g., see][]{2009grb..book.....V}
\be
\tau_{\rm \gamma\gamma,in}\simeq 10^{-1}\,\, R_{\rm dec, 17}\,\Gamma^{-1}_{1.5}\, n_{\rm \gamma,10.7}\,,
\ee
where  $R_{\rm dec}=2.7\times10^{17}\,{\rm cm}  \, n^{-\frac13}\, E^{\frac13}_{51.2}\,\Gamma^{-\frac23}_{1.5}$ is the radius at the deceleration phase and  $n_\gamma\simeq 5\times 10^{10}\,{\rm cm^{-3}}\, L_{\rm \gamma, 49}\,R^{-2}_{\rm dec, 17}\,\Gamma^{-1}_{1.5}\,\epsilon^{-1}_{\rm \gamma,3}$ is the density of the keV-energy photons associated with its photon luminosity.  Due to the fact that $\tau_{\rm \gamma\gamma,in}\ll 1$,  the intrinsic attenuation is not considered.\\ 
The minimum and the cooling electron Lorentz factors  are
{\small
\bary\label{elec_factor}
\gamma_{\rm m}&=& 4.3\times 10^2  \,\varepsilon_{\rm e,1.1} \,\Gamma_{1.5}, \cr
\gamma_{\rm c}&=& 8.4\times 10^5\, \left(\frac{1+z}{1.08} \right)\, [1+Y (\gamma_{\rm c})]^{-1} \,\varepsilon^{-1}_{B,-4}\,n^{-1}\Gamma^{-3}_{1.5}\,t^{-1}_3,\,\,\,\,\,\,\,\,\,
\eary
}
where $Y (\gamma_{\rm c})$ corresponds to the Compton parameter of the electrons with Lorentz factor $\gamma_c$ \citep{2010ApJ...712.1232W}. The value of $Y (\gamma_{\rm c})$ is given by 

{\footnotesize
\bary\label{Yc}
Y(\gamma_c)=   \frac{\eta \varepsilon_{\rm e}}{\varepsilon_{\rm B}[1+ Y(\gamma_c)]} \cases{ 
\left(\frac{\epsilon_{\rm KN}(\gamma_{\rm c})}{\epsilon_{\rm c}}\right)^{\frac{3-p}{2}}, \,  \epsilon^{\rm syn}_{\rm m} < \epsilon^{\rm syn}_{\rm KN}(\gamma_{\rm c}) < \epsilon^{\rm syn}_{\rm c}\,\,\,\,\,\,  \cr
 1.  \hspace{1.75 cm},\, \epsilon^{\rm syn}_{\rm c} < \epsilon^{\rm syn}_{\rm KN}(\gamma_{\rm c}), \cr
 }
 \eary 
}
with $\eta=\left(\frac{\gamma_{\rm c}}{\gamma_{\rm m}}\right)^{2-p}$ and  $\epsilon^{\rm syn}_{\rm KN}(\gamma_{\rm c})=\frac{\Gamma\,m_{\rm e}c^2 }{\gamma_{\rm c}}$  \citep{2009ApJ...703..675N, 2010ApJ...712.1232W}.   Given the synchrotron spectral breaks $\epsilon^{\rm syn}_{\rm m}=2.9\,{\rm eV}$,  $\epsilon^{\rm syn}_{\rm c}=0.1\,{\rm MeV}$ and $\epsilon^{\rm syn}_{\rm KN}(\gamma_{\rm c}) \simeq12.8\,{\rm eV}$ the Compton parameter lies in the range of $\epsilon^{\rm syn}_{\rm m} < \epsilon^{\rm syn}_{\rm KN}(\gamma_{\rm c})< \epsilon^{\rm syn}_{\rm c}$. Solving the eq. \ref{Yc}, the value of $Y(\gamma_c)$ becomes $\sim 0.8$.\\
Given the electron Lorentz factors (eq. \ref{elec_factor}) and the  synchrotron spectral breaks \citep{1998ApJ...497L..17S},  the spectral breaks  and the maximum flux for SSC emission are
{\small
\bary\label{ssc_br-h}
\epsilon^{\rm ssc}_{\rm m}&\simeq& 0.1\,{\rm MeV}\,  \left(\frac{1+z}{1.08} \right)^{\frac54}\,\varepsilon_{e,-1.1}^{4}\,\varepsilon_{B,-4}^{\frac12}\,n^{-\frac14}\,E^{\frac34}_{51.2}\,t^{-\frac94}_3,\cr
\epsilon^{\rm ssc}_{\rm c}&\simeq& 6.1\,10^{5}\,{\rm TeV} \, \left(\frac{1+z}{1.08} \right)^{-\frac34}\,\left(\frac{1+Y_{\rm Th}}{1.8} \right)^{-4}\,\varepsilon_{B,-4}^{-\frac72}\,n^{-\frac94}\,E^{-\frac54}_{51.2},\cr
&&\hspace{6.2cm}\times\,t^{-\frac14}_3\,,\cr
F^{\rm ssc}_{\rm max}&\simeq& 3.2\times 10^{-5}\,{\rm mJy} \,\left(\frac{1+z}{1.08} \right)^{\frac34}\,\varepsilon_{B,-4}^{\frac12}\,n^{\frac54}\,d^{-2}_{z,27}\,E^{\frac54}_{51.2}\, t^{\frac14}_3.
\eary
}
The cooling spectral break is too large in comparison with the Fermi LAT energy.    The SSC light curves in the fast (slow)-cooling regime  are given by \citep{2019ApJ...883..162F}
{\footnotesize
\begin{eqnarray}
\label{ssc_ism} 
F^{\rm ssc}_{\nu}\propto \cases{ 
t^{\frac{1}{8}} \epsilon_{\gamma}^{-\frac12}(t^{-\frac{9p-11}{8}} \epsilon_{\gamma,11}^{-\frac{p-1}{2}}), \hspace{0.4cm} \epsilon^{\rm ssc}_{\rm c} (\epsilon^{\rm ssc}_{\rm m})<\epsilon_\gamma<\epsilon^{\rm ssc}_{\rm m} (\epsilon^{\rm ssc}_{\rm c}),\hspace{.25cm}\cr
t^{-\frac{9p-10}{8}}\,\epsilon_{\gamma}^{-\frac{p}{2}},\,\,\,\, \hspace{1.5cm}  \{\epsilon^{\rm ssc}_{\rm c},\epsilon^{\rm ssc}_{\rm m}\}<\epsilon_\gamma\,, \cr
}
\end{eqnarray}
}
The KN suppression effect is taken into account in the SSC model  because of the reduction of the emissivity in comparison with the Compton regime.  The break energy in the KN regime becomes
{
\small
\bary
\epsilon^{\rm KN}_{\rm c}&\simeq& 14.3\,{\rm TeV}\, \left(\frac{1+z}{1.08} \right)^{-\frac34}\, \left(\frac{1+Y_{\rm Th}}{1.8} \right)^{-1} \,\varepsilon_{B,-4}^{-1}\,n^{-\frac34}\,E^{-\frac14}_{51.2}\,\cr
&&\hspace{5.9cm}\times\, t^{-\frac14}_3.
\eary
}
\subsection{H.E.S.S. detection with Fermi LAT upper limits}
%
From the spectral breaks of synchrotron ($\epsilon^{\rm syn}_{\rm m}\simeq2.9\,{\rm eV}$ and $\epsilon^{\rm syn}_{\rm c}\simeq 0.1\,{\rm MeV}$)    and  SSC ($\epsilon^{\rm ssc}_{\rm m}\simeq 0.1\,{\rm MeV}$ and  $\epsilon^{\rm ssc}_{\rm c}\simeq 6.1\times 10^{5}\,{\rm TeV}$) emission, one can see that  at 100 MeV the synchrotron emission lies in the range $\epsilon^{\rm syn}_{\rm m}<\epsilon^{\rm syn}_{\rm c}  < \epsilon_\gamma$ and  the SSC emission in the range  $\epsilon^{\rm ssc}_{\rm m} <\epsilon_\gamma<\epsilon^{\rm ssc}_{\rm c}$.   It is important to consider electrons that have electron Lorentz factors $\gamma^*_{\rm e}$ and radiate synchrotron photons at 100 MeV.  Using the electron Lorentz factor that produces synchrotron photons at the LAT regime, it is possible to obtain the critical energy $\epsilon^{\rm syn}_{\rm KN}(\gamma^*_{\rm e})=\frac{\Gamma\,m_{\rm e}c^2 }{\gamma^*_{\rm e}}\simeq 1.4\,{\rm eV}$ \citep{2009ApJ...703..675N}.   In this case,  the Compton parameter becomes $Y(\gamma^*_{\rm e})=Y(\gamma_{\rm c})\, \left(\frac{\epsilon^{\rm syn}_{\rm KN}(\gamma^*_{\rm e})}{\epsilon^{\rm syn}_{\rm KN}(\gamma_{\rm c})}\right)^{\frac{3-p}{2}} \simeq 0.7$ which corresponds to the range $\epsilon^{\rm syn}_{\rm KN}(\gamma^*_{\rm e})<\epsilon^{\rm syn}_{\rm m}< \epsilon^{\rm syn}_{\rm KN}(\gamma_{\rm c})< \epsilon^{\rm syn}_{\rm c} $ \citep{2009ApJ...703..675N, 2010ApJ...712.1232W,  2015MNRAS.454.1073B}.\\
%
%
Therefore, the ratio of synchrotron and SSC fluxes at 100 MeV becomes
{\small
\bary\label{ism}
\frac{F^{\rm syn}_\nu}{F^{\rm ssc}_\nu} &\sim& 5.1 \left(\frac{1+z}{1.08} \right)^{-\frac14}\,\left(\frac{1+Y(\gamma^*_{\rm e})}{1.7} \right)^{-1}\, \varepsilon^{1-p}_{e,-1.1} \varepsilon^{-\frac34}_{B,-4}n^{-\frac76}\, E^{-\frac{1}{12}}_{51.2}\cr
&& \hspace{3.cm} \,\Gamma^{\frac53-p}_{1.5}t^{-\frac14}_3\,\left( \frac{\epsilon_\gamma}{100\,{\rm MeV}}\right)^{-\frac12}\,,
\eary
}
The ratio of the SSC to synchrotron luminosity is approximately given by $Y(\gamma_{\rm c})\equiv\frac{L^{\rm ssc}_\nu}{L^{\rm syn}_\nu}=\frac{U_{\rm syn}[\epsilon_\gamma <  \epsilon^{\rm syn}_{\rm KN}(\gamma_{\rm c})]}{U_{\rm B}}\simeq 0.8$ \citep{2010ApJ...712.1232W}. The terms $U_{\rm B}$ and $U_{\rm syn}[\epsilon_\gamma <  \epsilon^{\rm syn}_{\rm KN}   (\gamma_{\rm c})]$ are the energy densities of the magnetic field and the synchrotron photons with energy below $\epsilon^{\rm syn}_{\rm KN}(\gamma_{\rm c})$.
%
%
%
%
%
%
%
\\
The right-hand  panel  in Figure \ref{figure5}  shows  the SSC FS emission estimated at 100 MeV (the dashed magenta curve) and 80 GeV (the dotted-dashed green curve).  The attenuation produced by the extragalactic background light (EBL) absorption is accounted for using the model presented in \cite{2017A&A...603A..34F}.  The intrinsic attenuation by $e^\pm$ pair production is not taken into account  because during the deceleration phase it is not significant.     In order to verify the results of our model with the observations at high and very-high energies,  the Fermi-LAT upper limits at 100 MeV and the sensitivity of H.E.S.S. at 80 GeV are shown. This panel shows that the SSC emission can reproduce  the observations of H.E.S.S. and Fermi LAT (i.e., it can be detected  in the H.E.S.S. observatory without being detected by the Fermi LAT instrument).\\ 
\subsection{Why GRB 190829A was detectable at VHE}
As follows we enumerate the reasons why this burst could have been detected.\\
\subparagraph{i) A burst located at very low redshift with intermediate luminosity.}  Depending on the redshift and the photon energy, VHE flux from a generic source begins to be attenuated  due to $e^\pm$ pair creation with EBL photons \citep{1966PhRvL..16..252G}.  This attenuation can be measured through $\exp[-\tau_{\gamma\gamma} (z)]$ with $\tau_{\gamma\gamma}(z)$ the opacity.  Given the distance of GRB 190829A with a low-redshift of  $z\simeq0.078$,  the VHE flux at 100 GeV  and 1 TeV  is attenuated by a factor of $0.99$  and $0.44$ \citep{2017A&A...603A..34F}, respectively.  On the other hand, with the best-fit parameters found after describing GRB 190829A, the SSC flux evolves  in the range $\epsilon^{\rm ssc}_{\rm m}<\epsilon_\gamma< \epsilon^{\rm ssc}_{\rm c}$ and consequently, the SSC flux varies as  $\propto E^{1.1}$.  Therefore, the very small attenuation factor together with an intermediate luminosity allowed that GRB 190829A could have been  detected by the H.E.S.S. observatory.  It is worth noting that although no imaging atmospheric Cherenkov telescope (IACT) was observing GRB 130702A,  an intermediate luminosity burst with $z\lesssim0.2$, Fermi LAT detected a GeV-photon associated with this burst (see section \ref{GRB130702A}). \\

\subparagraph{ii) The VHE emission was released during the deceleration phase.}  The high- and very-high-energy photons have been observed during the prompt and afterglow in tens of bursts \citep{2019ApJ...878...52A}.  Different studies  of multi-wavelength observations  have yielded results about the places where VHE flux is originated \citep[e.g. internal and external shocks; ][]{2015PhR...561....1K}. In accordance with our model and the best-fit values found,  the VHE emission reported by H.E.S.S. was created during the deceleration phase with the intrinsic attenuation due to $\gamma\gamma$ interactions much less than unity. In this case, the intrinsic attenuation did not decrease the observed SSC emission.\\
\subparagraph{iii) A favorable set of parameters.} The set of the best-fit parameters as found for GRB 190829A made  more favorable its detection.   For instance, with the best-fit parameters   the SSC flux evolving in the second PL segment of slow-cooling regime increases as the circumburst density ($\propto n^{1.1}$) and electron equipartition parameter ($\propto \varepsilon_{\rm e}^{2.3}$)  increase.   Higher values of these parameters make SSC emission more favorable to be detected. \\
\subparagraph{iv) The KN regime is much above hundreds of GeVs.}  The electron distribution up-scatters synchrotron FS photons at the KN regime.   The SSC emission evolving in the KN regime  is drastically attenuated. For GRB 190829A,  the SSC flux  in the range of the H.E.S.S. is much below than the KN regime  ($\simeq$ 14.3 TeV). It allowed that  the SSC flux was not attenuated and hence, was detected by the H.E.S.S. telescopes.\\
\subparagraph{v) A quick location of this burst.}  To observe photons  at hundreds of GeVs by IACTs has been truly a difficult challenge  since these take longer in locating the burst  than the duration of the main emission  and the early afterglow. In spite of numerous attempts, two observations, GRB 180720B \citep{2019Natur.575..464A} and GRB 190114C \citep{2019Natur.575..459A}, have been possible and many upper VHE limits  have been derived by these telescopes \citep[e.g. see,][]{Albert_MAGIC, Aleksi_GRB090102,  Aharonian_2009HESS, Aharonian_GRB060602B, HESS_GRB100621A, Acciari_VERITAS,  Bartoli_ARGO, GRB150323A_Abeysekara}.   We argue that the conditions to quickly pinpoint  the early afterglow of GRB 190829A by the H.E.S.S. telescope  made the VHE flux detection possible.
\subsection{Analysis of our SSC model for other VHE observatories}
\subparagraph{CTA Telescopes.}  \cite{2013APh....43..348F}  and \cite{2016CRPhy..17..617P} presented and discussed the sensitivity to transient sources of the  Cherenkov Telescope Array (CTA) telescopes for distinct energy thresholds. At $10^3$ s, the CTA sensitivity for an energy threshold of 75 GeVs is $\sim2.6\times 10^{-10}$ mJy. In order to compare with our SSC model,  the corresponding SSC flux, at $t=10^3$ s  and $\epsilon_\gamma=75\, {\rm GeV}$,  is  $\simeq 9.0\times 10^{-9}\,{\rm mJy}$.    Hence,  we conclude that GRB 190829A would have been detected by CTA if this would have been working.  GRBs with similar characteristics of GRB 190829A are potential candidates to be detected \citep[e.g. CTA;][]{2013APh....43..348F}. \\ 
\subparagraph{Magic Telescopes.}  \cite{2008ApJ...687L...5T} reported the MAGIC GRB sensitivity as a function of time at 100 GeV. At $10^3$ s, the MAGIC sensitivity becomes $\sim 10^{-10}\,{\rm mJy}$.  The corresponding SSC flux, at $t=10^3$ s  and $\epsilon_\gamma=100\,{\rm GeV}$, would be $\simeq 7.5\times 10^{-9}\,{\rm mJy}$.  Therefore, we conclude that if this burst would have been early located by the MAGIC telescopes, GRB 190829A would have been detected.\\
\subparagraph{HAWC Observatory.}   The High Altitude Water Cherenkov (HAWC) collaboration reported the GRB  sensitivity as a function of time for distinct zenith angles in the energy range of (0.1 - 1) TeV \citep{2019arXiv190806122M}. Considering the zenith angle with respect of GRB 190829A ($\theta_{\rm zenith}=30^\circ$) at $10^3\,{\rm s}$, the  flux sensitivity is around $\simeq 10^{-8}\,{\rm erg\,cm^{-2}\,s^{-1}}$. Taking into account the EBL absorption effect in the (0.1 - 1) TeV energy range, the SSC flux     would be  $\simeq 10^{-9}\,{\rm erg\,cm^{-2}\,s^{-1}}$.  We conclude that GRB 190829A could not have been detected by the HAWC gamma-ray observatory, even if this burst would have occurred during the first hours in the HAWC field of view.\\
\subsection{Generalization to the closest ilGRBs  ($z\lesssim 0.2$)}\label{GRB130702A}
To date, there is only one confirmed ilGRB/SN detected with $z\lesssim 0.2$; GRB 130702A.   GRB 130702A classified as an intermediate-luminosity burst and associated to a broad-line, type Ic supernovae SN2013dx \citep{2015A&A...577A.116D}, was detected in different wavelengths ranging from radio to high-energy gamma-rays. The Fermi GBM instrument  triggered on GRB 130702A at 2013 July 02 00:05:23.079 UTC. The Fermi LAT instrument detected photons from this burst above $>$100 MeV  within 2200 s.  The duration of the main emission in the (50-300) keV energy range was $T_{90}=59\,{\rm s}$ and the isotropic energy reported was $6.4^{+1.3}_{-1.0}\times 10^{50}\,{\rm erg}$ for a redshift of  z=0.145 \citep{2016ApJ...818...79T}. Details of the data analysis and the afterglow observations are reported in  \cite{2015A&A...577A.116D} and  \cite{2016ApJ...818...79T}.\\
\subsubsection{Fermi-LAT analysis and synchrotron limit}
\begin{figure}[h!]
{\centering
\resizebox*{0.48\textwidth}{0.33\textheight}
{\includegraphics{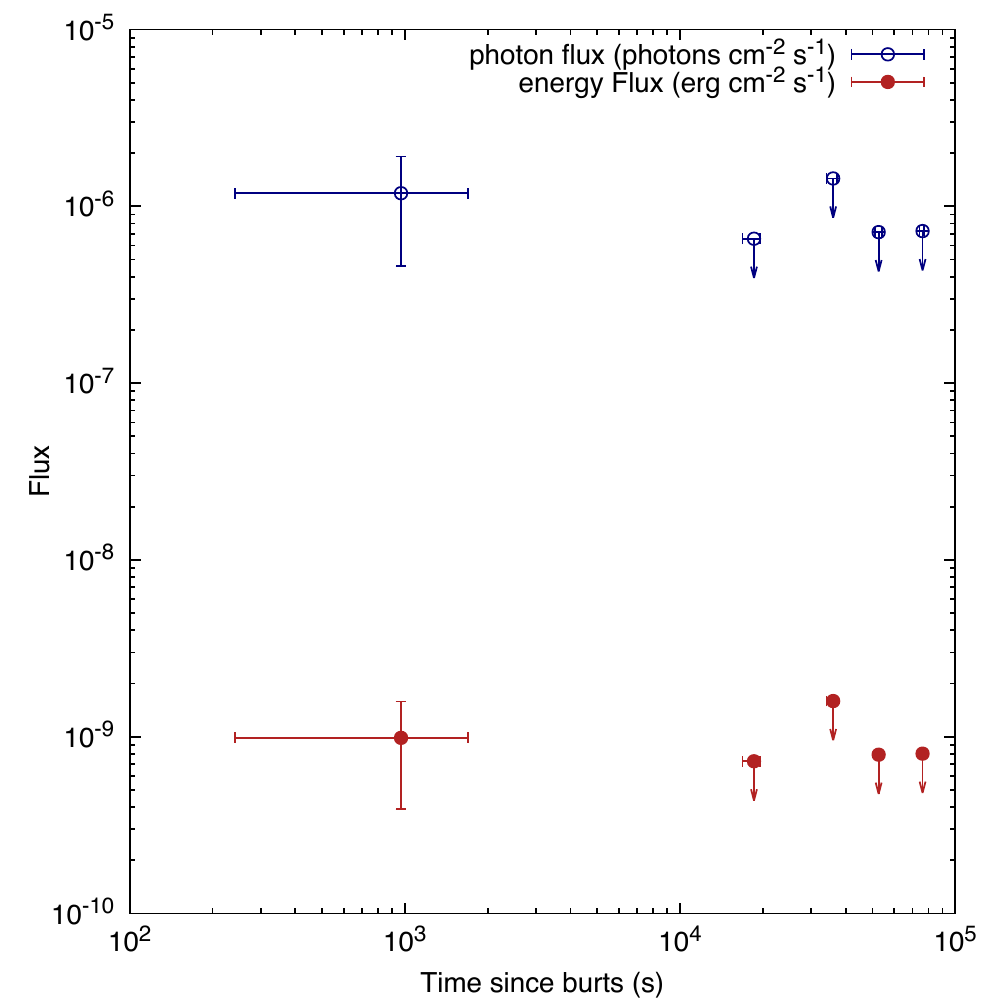}}
\resizebox*{0.48\textwidth}{0.33\textheight}
{\includegraphics{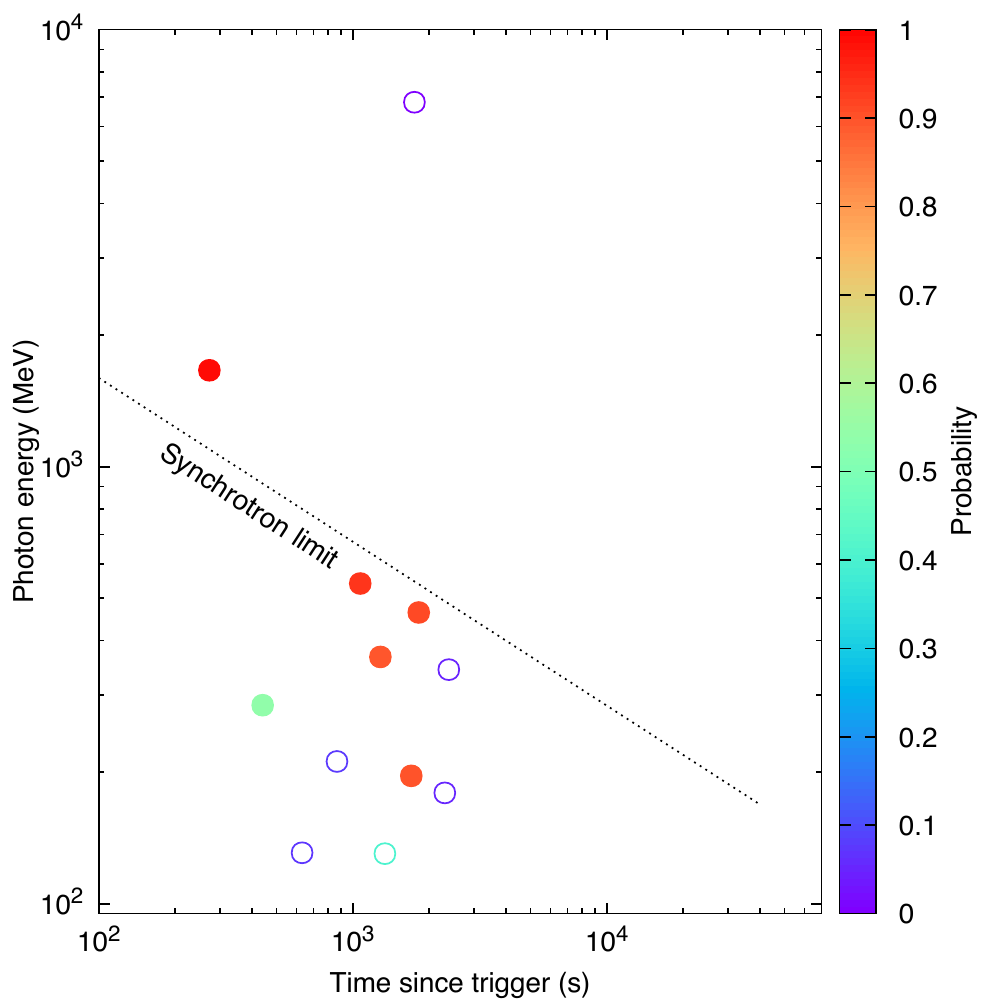}}
}
\caption{The left-hand panel shows the Fermi-LAT energy flux (blue) and photon flux (red) light curves obtained between 0.1 and 100 GeV. The right-hand panel shows all the photons with energies $> 100$ MeV and their respective probabilities  to be associated to  GRB 130702A. The dotted line corresponds to the synchrotron limit.}
\label{figure6}
\end{figure}

The left-hand panel in Figure \ref{figure6} displays the Fermi LAT energy flux (blue) and photon flux (red) light curves and upper limits obtained between 0.1 and 100 GeV.   This panel shows that the flux at $\sim 10^3\,{\rm s}$ is slightly above the upper limits at $> 10^{4}\,{\rm s}$.   The right-hand panel in Figure \ref{figure6} shows all the photons with energies above $> 100$ MeV as a function of probabilities to be associated to GRB 130702A. The data files used for this analysis are given at the data website.\footnote{https://fermi.gsfc.nasa.gov/cgi-bin/ssc/LAT/LATDataQuery.cgi} Details of the fermi tools and the procedure to analyze the Fermi LAT data are presented in \cite{2019ApJ...885...29F}.   Several features can be noted:  i) The Fermi LAT detected three high-energy photons with probabilities $>$ 90\% of 1661, 540 and 464 MeV detected at 272, 1070 and 1818 s, respectively, after the trigger time,  ii) the highest energy photon of 1661 MeV corresponded to the first  photon observed at $272\,{\rm s}$  and iii) this burst displayed 5 photons above $>$ 100 MeV with a probability less than 10\% (the highest energy photon was 7 GeV).\\
Taking into account the value of circumburst density $n=1\,{\rm cm^{-3}}$, the total isotropic-equivalent energy $6.4^{+1.3}_{-1.0}\times 10^{50}\,{\rm erg}$  \citep{2015A&A...577A.116D}, the  redshift of  z=0.145 \citep{2016ApJ...818...79T}  and the efficiency to convert the kinetic energy into photons $0.2$ \citep{2015MNRAS.454.1073B},     we estimate and plot the evolution of the maximum energy photon radiated by synchrotron FS model, as shown in the right-hand panel in  Figure \ref{figure6}.  It is shown that the highest energy photon detected by the Fermi LAT cannot be interpreted in the synchrotron FS scenario, an additional mechanism is required  to interpret this photon. Therefore, as concluded for GRB 190829A an additional mechanism such SSC emission should be present during the FSs to explain this GeV energy photon. 
\vspace{0.5cm}
\subsubsection{Parameter space so that ilGRBs can be detected by the H.E.S.S. telescopes}
\begin{figure}[h!]
{\centering
\resizebox*{0.48\textwidth}{0.33\textheight}
{\includegraphics{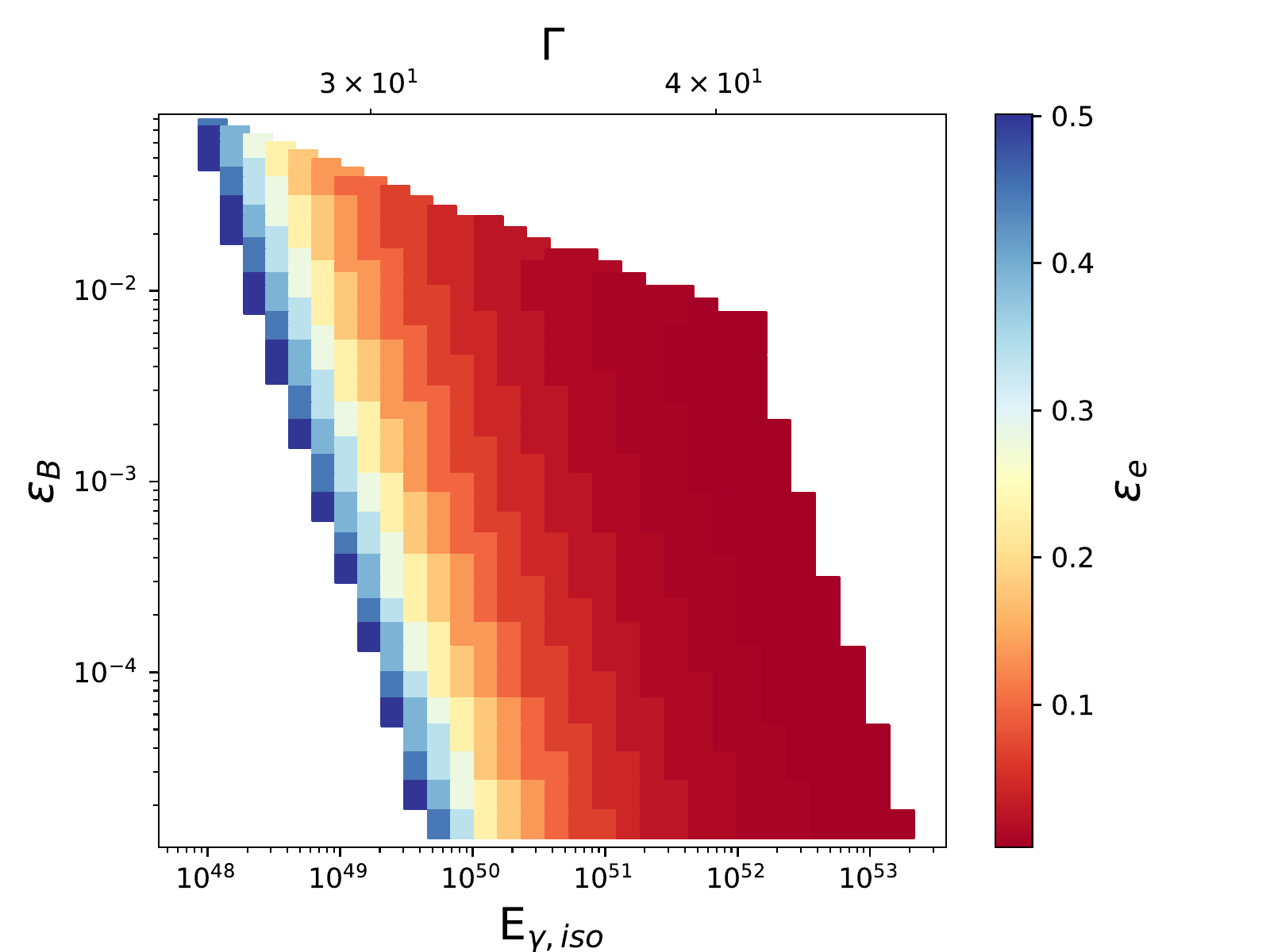}}
\resizebox*{0.48\textwidth}{0.33\textheight}
{\includegraphics{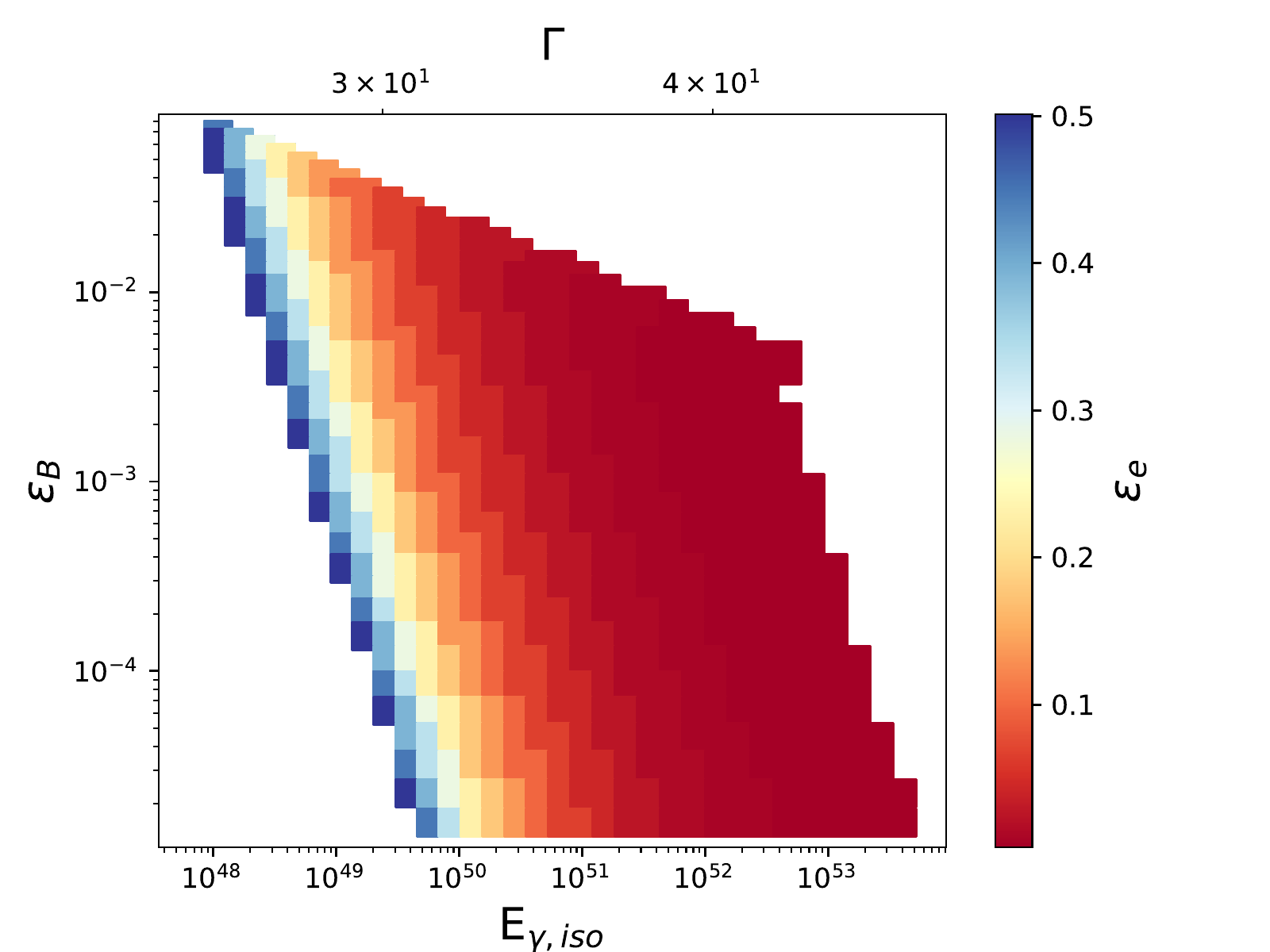}}
}
\caption{The left-hand panel shows the 4D parameter space of the microphysical parameters, isotropic energy and bulk Lorentz factor for which SSC flux from FS  is below (left)   and above 5 times (right) the LAT sensitivity at 10 GeV and above the H.E.S.S. sensitivity at 80 GeV  \citep{2016CRPhy..17..617P}.  The upper ($\Gamma$) and the lower ($E_{\rm \gamma,iso}$)  X-axes are related through the deceleration time  of $10^3\,{\rm s}$ and the density of $1\,{\rm cm^{-3}}$.}
\label{figure7}
\end{figure}

Using the SSC model during the FSs (eqs. \ref{ssc_ism} and \ref{ssc_br-h}),  we compute the parameter space so that ilGRBs can be detected by the H.E.S.S. telescopes with the condition that the gamma-ray emission is below and slightly above (5 times)\footnote{ This value approximately corresponds to the difference between the upper limits derived in GRB 190829A and the observed flux in GRB 130702A.} the Fermi LAT sensitivity. In both cases, we consider the EBL model derived by  \cite{2017A&A...603A..34F},  a uniform-density medium of $n=1\,{\rm cm^{-3}}$, an electron spectral index of $p=2.1$, a redshift of  $z=0.2$ and a deceleration time of $10^3\,{\rm s}$.  
\paragraph{Below the Fermi LAT sensitivity.}
We plot the parameter space for which VHE gamma-ray emission can be detected in H.E.S.S. but not in the Fermi LAT.    The left-hand panel  in Figure \ref{figure7} shows the parameter space of the microphysical parameters, isotropic-equivalent energy and the bulk Lorentz factor for which SSC flux is below the Fermi LAT sensitivity at 10 GeV and above the H.E.S.S. sensitivity at 80 GeV  \citep{2016CRPhy..17..617P}.  The upper ($\Gamma$) and the lower ($E_{\rm \gamma,iso}$)  X-axes are related through the deceleration time  of $10^3\,{\rm s}$ and the density of $1\,{\rm cm^{-3}}$. In particular, this panel shows that for $E_{\rm \gamma, iso}\approx 10^{48}\,{\rm erg}$, the set of parameters that satisfy the conditions are  $\varepsilon_{\rm e}\gtrsim 0.3$, $\varepsilon_{\rm B}\gtrsim 10^{-2}$, $\Gamma\sim 25$, and for $E_{\rm \gamma, iso}\approx 10^{53}\,{\rm erg}$, the set of parameters become $\varepsilon_{\rm e}\lesssim 0.1$, $\varepsilon_{\rm B}\lesssim 10^{-5}$ and  $\Gamma\sim 45$. It shows that VHE photons cannot be expected from llGRBs with $\Gamma<10$.    The values of the initial Lorentz factor lies in the range of $25 \lesssim \Gamma \lesssim 45$. It is worth noting that if the deceleration time decreases about one-third of the assumed value,  the initial Lorentz factor would lie in the range of $40 \lesssim \Gamma \lesssim 70$.
\paragraph{Slightly above the Fermi LAT sensitivity}
We plot the parameter space for which VHE gamma-ray emission can be detected in H.E.S.S. and also slightly detectable in the Fermi LAT.   The right-hand panel  in Figure \ref{figure7} shows the parameter space of the microphysical parameters, isotropic-equivalent energy and the bulk Lorentz factor for which SSC flux is above the H.E.S.S. sensitivity at 80 GeV  \citep{2016CRPhy..17..617P} and  slightly above the Fermi LAT sensitivity at 10 GeV.  Again, the upper ($\Gamma$) and the lower ($E_{\rm \gamma,iso}$)  X-axes are related through the deceleration time  of $10^3\,{\rm s}$ and the density of $1\,{\rm cm^{-3}}$.   In particular, this panel shows that for $E_{\rm \gamma, iso}\approx 10^{48}\,{\rm erg}$, the set of parameters are  $\varepsilon_{\rm e}\gtrsim 0.5$, $\varepsilon_{\rm B}\gtrsim 5\times 10^{-2}$, $\Gamma\sim 28$, and for $E_{\rm \gamma, iso}\approx 5\times 10^{53}\,{\rm erg}$, the set of parameters become $\varepsilon_{\rm e}\lesssim 0.1$, $\varepsilon_{\rm B}\lesssim 10^{-5}$ and  $\Gamma\sim 48$. In this case,  the values of the initial Lorentz factor lies in the range of $28 \lesssim \Gamma \lesssim 48$.%
\section{Summary}
%
%
%
GRB 190829A, one of the closest bursts to Earth, was followed-up by a large number of satellites and observatories in several wavelengths that range from radio bands to hundreds of GeV gamma-rays.  Analysis of the prompt gamma-ray emission pointed to GRB 190819A as an intermediate luminosity burst ($10^{48.5}\,\lesssim L_{\rm iso}  \lesssim  10^{49.5} {\rm erg}$) and modelling the X-ray and optical light curves together with its SED indicate  that the outflow expands with an initial bulk Lorentz factor of $\Gamma\simeq 34$, which is high for a llGRB (e.g., $\Gamma\lesssim 10$) and low for a hlGRB (e.g., $\Gamma\gtrsim 100$). Thus,  GRB 190829A becomes the first intermediate-luminosity burst in being detected in the VHE gamma-ray band by an IACT, and, in turn,  the first  event what was not simultaneously observed by the Fermi LAT instrument.  It is worth noting that the value of bulk Lorentz factor is strongly constrained by the observation of the deceleration peak and that this value is also consistent with the required range for the Lorentz factor needed to produce VHE photons via SSC without also producing an SSC signal in LAT.  Our results indicate that no photons with energies above $\geq 100$ MeV can be associated to GRB 190829A.      We found that the ``plateau" phase can be interpreted as the fall-back accretion of a millisecond pulsar and the X-ray and optical observations are consistent with synchrotron forward-shock emission evolving between the characteristic and cooling spectral breaks during the early/coasting phase and late afterglow  in a uniform-density medium.    Using the best-fit parameters found after modelling the X-ray and optical light curves of GRB 190829A, we show that the VHE emission reported by the H.E.S.S. experiment  cannot be interpreted in the synchrotron FS scenario,  so that an additional mechanism should be present during the FSs to explain the multi-GeV photons. We interpret the energetic photons above the synchrotron limit in the SSC scenario during the FSs.   It is worth noting that high-energy photons detected by Fermi LAT below the synchrotron limit can be explained in terms of synchrotron FS scenario.  The detection of VHE emission above the synchrotron limit in GRB 190829A can be explained considering: i) the  very low redshift of this burst with an intermediate luminosity, ii) their origin during the deceleration phase, iii) the favorable set of parameters, iv) the KN regime much above hundreds of GeVs and finally, v) a quick location by H.E.S.S. telescope.\\
To date, there is only one confirmed ilGRB/SN detected with $z\lesssim 0.2$,  GRB 130702A. We have obtained the Fermi LAT light curve with its upper limits around the reported position of GRB 130702A and all photons with energies larger than $>$ 100 MeV. With a probability of $>$ 90\%, three high-energy photons of 1661, 540 and 464 MeV were detected during the afterglow phase at at 272, 1070 and 1818 s, respectively. It is shown that the highest energy photon cannot be interpreted in the synchrotron FS scenario, an additional mechanism is required  to interpret this photon. Therefore, as concluded for GRB 190829A an additional mechanism such as SSC should be present during the FSs to explain this GeV energy photon.\\ 
Considering that synchrotron FS model is not sufficient to explain the high-energy and VHE photons in the two ilGRBs,  we finally compute the parameter space so that SSC flux originated in these object could be detected by the H.E.S.S. telescopes. We show that low-redshift bursts with intermediate luminosity are potential candidates to be detected in very-high energies.\\
\acknowledgements
NF  acknowledges  financial  support  from UNAM-DGAPA-PAPIIT  through  grant  IA102019.  RBD  acknowledges support  from the National Science Foundation under grant 1816694.
%
%
%
%

%

%
\clearpage

\clearpage

\begin{table}
\centering
\caption{Spectral analysis using the GBM data}\label{table1:gbm_analysis}
\begin{tabular}{ l c c c c c c}
 \hline
 \scriptsize{Time Interval (s)} & \scriptsize{$\alpha_{\rm Band}$}& \scriptsize{$\beta_{\rm Band}$} &\scriptsize{$E_{\rm peak}$ (keV)}  &\scriptsize{$E_{\rm iso}\,(\rm erg)$}  &\scriptsize{$F\,({\rm erg\,cm^{-2}\,s^{-1}})$}  \\
 \hline 
 \hline\\
Initial pulse$^a$\\ 
\scriptsize{[-2.0 ; 0.0]}  & \scriptsize{-$0.10\pm0.02$}  &  \scriptsize{-$1.32\pm0.10$} & \scriptsize{$11.4\pm1.7$} &   \scriptsize{$(3.0\pm0.2) \times 10^{49}$}  &   \scriptsize{$(2.22\pm 0.31) \times 10^{-7}$}   \\
\scriptsize{[0.0 ;  2.0]}  & \scriptsize{-$0.72\pm0.45$}  &  \scriptsize{-$1.80\pm0.08$} & \scriptsize{$62.3\pm25.1$} &   \scriptsize{$(3.7\pm0.2) \times 10^{49}$}  &   \scriptsize{$(6.41\pm 0.39) \times 10^{-7}$}  \\
\scriptsize{[2.0 ; 4.0]}  & \scriptsize{-$1.15\pm0.25$}  &  \scriptsize{-$2.53\pm0.41$} & \scriptsize{$59.0\pm12.4$} &   \scriptsize{$(1.1\pm 0.1)\times 10^{49}$}   &   \scriptsize{$(3.05\pm 0.34) \times 10^{-7}$} \\
\scriptsize{[4.0 ; 6.0]}  & \scriptsize{-$0.10\pm0.02$}  &  \scriptsize{-$2.15\pm0.13$} & \scriptsize{$20.5\pm4.3$} &   \scriptsize{$(7.3\pm 0.4) \times 10^{48}$}  &   \scriptsize{$(1.79\pm 0.24) \times 10^{-7}$}  \\
\scriptsize{[6.0 ; 8.0]}  & \scriptsize{-$0.10\pm0.02$}  &  \scriptsize{-$2.50\pm0.38$} & \scriptsize{$15.1\pm3.9$} &   \scriptsize{$(2.9\pm 0.2) \times 10^{48}$}   &   \scriptsize{$(7.30\pm 0.81) \times 10^{-8}$} \\
\scriptsize{[8.0 ; 10.0]}  & \scriptsize{-$0.10\pm0.02$}  &  \scriptsize{-$2.38\pm0.32$} & \scriptsize{$11.4\pm1.7$} &   \scriptsize{$(2.0\pm0.1)\times 10^{48}$}   &   \scriptsize{$(4.69\pm 0.16) \times 10^{-8}$} \\
\scriptsize{[10.0 ; 12.0]}  & \scriptsize{-$0.10\pm0.02$}  &  \scriptsize{-$2.50\pm0.38$} & \scriptsize{$15.2\pm10.4$} &   \scriptsize{$(1.0\pm0.1)\times 10^{48}$}  &   \scriptsize{$(2.61\pm 0.77) \times 10^{-8}$}  \\
\hline\\
Brighter peak$^b$\\ 
\scriptsize{[46.0 ; 48.0]}  & \scriptsize{-$1.11\pm0.17$}  &  \scriptsize{-$2.20\pm0.18$} & \scriptsize{$14.0\pm9.9$} &   \scriptsize{$(5.1\pm 0.1)\times 10^{48}$}  &   \scriptsize{$(1.09\pm 0.21) \times 10^{-7}$}   \\
\scriptsize{[48.0 ; 50.0]}  & \scriptsize{-$1.18\pm0.93$}  &  \scriptsize{-$2.46\pm0.03$} & \scriptsize{$12.0\pm2.3$} &   \scriptsize{$(4.7\pm 0.1)\times 10^{49}$}  &   \scriptsize{$(9.47\pm 0.03) \times 10^{-7}$}   \\
\scriptsize{[50.0 ; 52.0]}  & \scriptsize{-$1.11\pm1.12$}  &  \scriptsize{-$2.48\pm0.02$} & \scriptsize{$10.8\pm2.6$} &   \scriptsize{$(7.4\pm 0.1)\times 10^{49}$}  &   \scriptsize{$(1.47\pm 0.03) \times 10^{-6}$}   \\
\scriptsize{[52.0 ; 54.0]}  & \scriptsize{-$0.82\pm1.46$}  &  \scriptsize{-$2.50\pm0.02$} & \scriptsize{$11.1\pm2.0$} &   \scriptsize{$(6.5\pm 0.1)\times 10^{49}$}  &   \scriptsize{$(1.34\pm 0.02) \times 10^{-6}$}   \\
\scriptsize{[54.0 ; 56.0]}  & \scriptsize{-$1.19\pm1.22$}  &  \scriptsize{-$2.58\pm0.04$} & \scriptsize{$10.0\pm3.9$} &   \scriptsize{$(4.6\pm0.1)\times 10^{49}$}  &   \scriptsize{$(8.44\pm 0.21) \times 10^{-7}$}   \\
\scriptsize{[56.0 ; 58.0]}  & \scriptsize{-$1.11\pm0.17$}  &  \scriptsize{-$2.58\pm0.05$} & \scriptsize{$8.9\pm1.3$} &   \scriptsize{$(2.82\pm 0.03)\times 10^{49}$}  &   \scriptsize{$(4.92\pm 0.17) \times 10^{-7}$}   \\
\scriptsize{[58.0 ; 60.0]}  & \scriptsize{-$1.11\pm0.17$}  &  \scriptsize{-$2.64\pm0.09$} & \scriptsize{$9.7\pm1.6$} &   \scriptsize{$(1.43\pm0.02)\times 10^{49}$}  &   \scriptsize{$(2.54\pm 0.15) \times 10^{-7}$}   \\
\scriptsize{[60.0 ; 62.0]}  & \scriptsize{-$1.11\pm0.17$}  &  \scriptsize{-$2.66\pm0.13$} & \scriptsize{$7.1\pm4.2$} &   \scriptsize{$(8.4\pm 0.1)\times 10^{48}$}  &   \scriptsize{$(1.27\pm 0.13) \times 10^{-7}$}   \\
\scriptsize{[62.0 ; 64.0]}  & \scriptsize{-$1.11\pm0.17$}  &  \scriptsize{-$2.50\pm0.17$} & \scriptsize{$11.5\pm1.7$} &   \scriptsize{$(5.2\pm 0.1)\times 10^{48}$}  &   \scriptsize{$(1.05\pm 0.16) \times 10^{-7}$}   \\
\scriptsize{[64.0 ; 66.0]}  & \scriptsize{-$1.11\pm0.17$}  &  \scriptsize{-$2.29\pm0.29$} & \scriptsize{$11.5\pm1.7$} &   \scriptsize{$(2.52\pm 0.03)\times 10^{48}$}  &   \scriptsize{$(5.20\pm 1.78) \times 10^{-8}$}   \\
\scriptsize{[66.0 ; 68.0]}  & \scriptsize{-$1.11\pm0.17$}  &  \scriptsize{-$2.55\pm0.38$} & \scriptsize{$20.6\pm10.8$} &   \scriptsize{$(1.81\pm 0.02)\times 10^{48}$}  &   \scriptsize{$(4.33\pm 0.94) \times 10^{-8}$}   \\
\hline
\end{tabular}
\begin{flushleft}
\scriptsize{ 
$^a$ The total isotropic-equivalent energy and the peak  energy correspond to $(9.151\pm 0.504)\times 10^{49}\,{\rm erg}$ and  $67.88\pm 23.34$ keV, respectively.\\
$^b$ The total isotropic-equivalent energy and the peak  energy correspond to $(2.967\pm 0.032)\times 10^{50}\,{\rm erg}$ and $11.47\pm 0.36$ keV, respectively.  \\
}
\end{flushleft}
\end{table}
\begin{table}
\centering \renewcommand{\arraystretch}{2}\addtolength{\tabcolsep}{3pt}
\caption{The best-fit parameters from the initial pulse and the brighter peak displayed in the GBM  light curve}\label{table2:fit_GBM}
\begin{tabular}{ c c c c c c c}
 \hline \hline
\scriptsize{Event} &\hspace{0.5cm}   \scriptsize{Period}  &\hspace{0.5cm}   \scriptsize{$\tau_1$}  & \hspace{0.5cm} \scriptsize{ $\tau_2$} & \hspace{0.5cm}  \scriptsize{$\alpha_{\rm \gamma}$}  & \hspace{0.5cm} \scriptsize{$t_0$} & \hspace{0.5cm} \scriptsize{ $\chi^2$/ndf} \\ 
\scriptsize{} & \hspace{0.5cm}  \scriptsize{(s)}   & \hspace{0.5cm}  \scriptsize{(s)}   & \hspace{0.5cm} \scriptsize{(s)}  &\hspace{0.5cm}\scriptsize{} & \hspace{0.5cm} \scriptsize{(s)} \\ 
\hline \hline
\scriptsize{Initial pulse}   	        & \hspace{0.5cm} \scriptsize{[-2.0 ;  10]}  &\hspace{0.5cm} \scriptsize{$2.31\pm 0.06$}       		&\hspace{0.5cm} \scriptsize{$0.39\pm 0.10$} &\hspace{0.5cm} \scriptsize{$-$}	&\hspace{0.5cm}  \scriptsize{$0.11\pm0.06$}&\hspace{0.5cm}  \scriptsize{$0.81$}\\	
\scriptsize{Brighter peak}   	        & \hspace{0.5cm} \scriptsize{[46 ; 68]}  &\hspace{0.5cm} \scriptsize{$32.2\pm7.2$}	&\hspace{0.5cm} \scriptsize{$-$} &\hspace{0.5cm} \scriptsize{$6.59\pm1.65 $}	&\hspace{0.5cm}  \scriptsize{$43.6\pm9.1$} &\hspace{0.5cm}  \scriptsize{$0.84$}\\
\hline \hline
\end{tabular}
\begin{flushleft}
\scriptsize{
}
\end{flushleft}
\end{table}
\begin{table}
\centering \renewcommand{\arraystretch}{2}\addtolength{\tabcolsep}{3pt} 
\caption{The best-fit parameters obtained  from the UVOT light curves. The theoretical values are estimated  for $p=2.15\pm0.17$.}\label{table4:optical}
\begin{tabular}{ c c c c c c c}
 \hline \hline
\scriptsize{Band} &\hspace{0.5cm}   \scriptsize{Rise index/Theory}  &\hspace{0.5cm}   \scriptsize{Peak time }  &\hspace{0.5cm}   \scriptsize{Decay index/Theory}  & \hspace{0.5cm}  \scriptsize{$\Delta t/t$}  & \hspace{0.5cm}  \scriptsize{$F_{\rm cont}$}    & \hspace{0.5cm} \scriptsize{ $\chi^2$/ndf} \\ 
\scriptsize{} & \hspace{0.5cm}  \scriptsize{ -($\alpha_{\rm O,r}$)}  & \hspace{0.5cm}  \scriptsize{$t_{\rm br}$ ($\times10^3$ s)}   &\hspace{0.5cm}  \scriptsize{ ($\alpha_{\rm O, d}$)} & \hspace{0.5cm}  \scriptsize{ } & \hspace{0.5cm}  \scriptsize{($\times 10^{-2}\,{\rm mJy}$)}   &\hspace{0.5cm} \\ 
\hline \hline
\scriptsize{V}	        & \hspace{0.5cm} \scriptsize{$3.28\pm0.43$ / $3.0$ } & \hspace{0.5cm} \scriptsize{$1.39\pm0.30$ }   &\hspace{0.5cm} \scriptsize{$1.19\pm0.30$ / $0.86\pm 0.13$}      & \hspace{0.5cm}  \scriptsize{$1.01\pm0.22$}  & \hspace{0.5cm}  \scriptsize{ $5.58\pm0.45$}        &\hspace{0.5cm}  \scriptsize{$0.87$}             \\
\scriptsize{B}     	        & \hspace{0.5cm} \scriptsize{$2.87\pm0.41$ / $3.0$ } & \hspace{0.5cm} \scriptsize{$1.41\pm0.32$}  &\hspace{0.5cm} \scriptsize{$1.13\pm0.38$ / $0.86\pm 0.13$}	& \hspace{0.5cm}  \scriptsize{$0.68\pm0.21$} & \hspace{0.5cm}  \scriptsize{ $2.56\pm0.34$}  	&\hspace{0.5cm} \scriptsize{$1.25$} 								                     \\
\scriptsize{White}   	        &\hspace{0.5cm} \scriptsize{$3.08\pm0.31$ / $3.0$} & \hspace{0.5cm} \scriptsize{$1.38\pm0.29$ }  &\hspace{0.5cm} \scriptsize{$1.03\pm0.25$ / $0.86\pm 0.13$}	& \hspace{0.5cm}  \scriptsize{$0.71\pm0.23$}  & \hspace{0.5cm}  \scriptsize{$1.31\pm0.25$} 	&\hspace{0.5cm}  \scriptsize{$1.71$}							                     \\
\scriptsize{U}   	        & \hspace{0.5cm} \scriptsize{$2.80\pm0.32$ / $3.0$}  &  \hspace{0.5cm} \scriptsize{$1.25\pm0.39$ }  &\hspace{0.5cm} \scriptsize{$0.76\pm0.22$ / $0.86\pm 0.13$}	& \hspace{0.5cm}  \scriptsize{$0.73\pm0.23$} & \hspace{0.5cm}  \scriptsize{$0.93\pm0.09$}  	&\hspace{0.5cm}   \scriptsize{$0.81$}									                     \\
\\
\hline \hline
\end{tabular}
\end{table}
\begin{table}
\centering \renewcommand{\arraystretch}{2}\addtolength{\tabcolsep}{3pt}
\caption{The best-fit parameter obtained from different epochs identified in the Swift (BAT + XRT) light curve. The theoretical values are estimated  for $p=2.15\pm0.17$.} \label{table3:X_ray}
\begin{tabular}{c c c  c c c}
 \hline \hline
\scriptsize{Epochs} &\hspace{0.5cm}   \scriptsize{Period}  &\hspace{0.5cm}   \scriptsize{Index/Theory}  &\hspace{0.5cm}   \scriptsize{Peak time (s)}  & \hspace{0.5cm}  \scriptsize{$\Delta t/t$}    & \hspace{0.5cm} \scriptsize{ $\chi^2$/ndf} \\ 
\scriptsize{} & \hspace{0.5cm}  & \hspace{0.5cm}  \scriptsize{}   &   \hspace{0.5cm}  \scriptsize{}  & \\ 
\hline \hline
\scriptsize{0}   	        & \hspace{0.5cm} \scriptsize{$52 - 62$ s}  &\hspace{0.5cm} \scriptsize{$2.72\pm0.28$/$3.1\pm 0.15$}	&\hspace{0.5cm} \scriptsize{$-$} &\hspace{0.5cm} \scriptsize{$-$}	&\hspace{0.5cm}  \scriptsize{$0.75$}\\
\scriptsize{}   	        & \hspace{0.5cm} \scriptsize{}  &\hspace{0.5cm} \scriptsize{}	&\hspace{0.5cm} \scriptsize{} &\hspace{0.5cm} \scriptsize{}	&\hspace{0.5cm}  \scriptsize{}\\ \cdashline{1-6}
\scriptsize{I}   	        & \hspace{0.5cm} \scriptsize{$82 - 215$ s}  &\hspace{0.5cm} \scriptsize{$3.53\pm0.70$ / $3.1\pm 0.15$}	&\hspace{0.5cm} \scriptsize{$-$} &\hspace{0.5cm} \scriptsize{$-$}	&\hspace{0.5cm}  \scriptsize{$0.81$}\\	
\scriptsize{}   	        & \hspace{0.5cm} \scriptsize{}  &\hspace{0.5cm} \scriptsize{$$}	&\hspace{0.5cm} \scriptsize{$$} &\hspace{0.5cm} \scriptsize{$$}	&\hspace{0.5cm}  \scriptsize{$$}\\ \cdashline{1-6}	
\scriptsize{II}   	        & \hspace{0.5cm} \scriptsize{$215 - 700$ s}  &\hspace{0.5cm} \scriptsize{$0.06\pm0.01$ / 0}	&\hspace{0.5cm} \scriptsize{$-$} &\hspace{0.5cm} \scriptsize{$-$}	&\hspace{0.5cm}  \scriptsize{$0.84$}\\
\scriptsize{}   	        & \hspace{0.5cm} \scriptsize{}  &\hspace{0.5cm} \scriptsize{$$}	&\hspace{0.5cm} \scriptsize{$$} &\hspace{0.5cm} \scriptsize{$$}	&\hspace{0.5cm}  \scriptsize{$$}\\ \cdashline{1-6}

\scriptsize{III}   	        & \hspace{0.5cm} \scriptsize{$700 - 1.4^{+0.17}_{-0.15} \times 10^5$ s}  &\hspace{0.5cm} \scriptsize{$-(3.12\pm0.94)$ / $-3.0$} &\hspace{0.5cm} \scriptsize{$1.4 \times 10^3$}	&\hspace{0.5cm} \scriptsize{$0.75\pm0.24$}	&\hspace{0.5cm}  \scriptsize{$0.83$}\\
\scriptsize{}   	        & \hspace{0.5cm} \scriptsize{}  &\hspace{0.5cm} \scriptsize{$1.03\pm0.12$ / $0.86\pm0.13$} &		&\hspace{0.5cm} \scriptsize{}&\hspace{0.5cm}  \scriptsize{}\\\cdashline{1-6}
\scriptsize{IV$^a$}   	        & \hspace{0.5cm} \scriptsize{$\geq 1.4^{+0.17}_{-0.15} \times 10^5$ s}  &\hspace{0.5cm} \scriptsize{$1.23\pm0.04$ / $1.11\pm0.13$}	&\hspace{0.5cm} \scriptsize{$-$}&\hspace{0.5cm} \scriptsize{$-$}	&\hspace{0.5cm}  \scriptsize{$0.91$}\\
\\\cdashline{1-6}
\hline
\end{tabular}
\begin{flushleft}
\scriptsize{
$^a$ These values are taken from the Swift analysis.\\
}
\end{flushleft}
\end{table}

\begin{table}
\centering \renewcommand{\arraystretch}{2}\addtolength{\tabcolsep}{3pt}
\caption{The best-fit parameters of the multi-peaks obtained from the XRT light curve at 10 ${\rm keV}$.} \label{table4:peaks}
\begin{tabular}{c c c  c c}
 \hline \hline
\scriptsize{Peaks}   &\hspace{0.5cm}   \scriptsize{Index}                       &\hspace{0.5cm}   \scriptsize{Peak time }                                & \hspace{0.5cm}  \scriptsize{$\Delta t/t$}    & \hspace{0.5cm} \scriptsize{ $\chi^2$/ndf} \\ 
\scriptsize{}               & \hspace{0.5cm}  \scriptsize{($\alpha_{\rm X}$)}   &  \hspace{0.5cm}  \scriptsize{(s)}  &    & \\ 
\hline \hline
\scriptsize{a}   	          &\hspace{0.5cm} \scriptsize{$-(22.9\pm6.2)$}	      &  \hspace{0.5cm} \scriptsize{$97.5$}                                          &\hspace{0.5cm} \scriptsize{$0.15$}	&\hspace{0.5cm}  \scriptsize{$1.7$}\\
	                           &\hspace{0.5cm} \scriptsize{$3.5\pm1.3$}	      &  \hspace{0.5cm} \scriptsize{$$}                                          &\hspace{0.5cm} \scriptsize{$$}	&\hspace{0.5cm}  \scriptsize{}\\ \cdashline{1-5}
\scriptsize{b}   	          &\hspace{0.5cm} \scriptsize{$-(11.0\pm0.9)$}	      &  \hspace{0.5cm} \scriptsize{$133.0$}                                          &\hspace{0.5cm} \scriptsize{$0.14$}	&\hspace{0.5cm}  \scriptsize{$1.1$}\\
	                           &\hspace{0.5cm} \scriptsize{$13.05\pm1.5$}	      &  \hspace{0.5cm} \scriptsize{$$}                                          &\hspace{0.5cm} \scriptsize{$$}	&\hspace{0.5cm}  \scriptsize{}\\ \cdashline{1-5}
\scriptsize{c}   	          &\hspace{0.5cm} \scriptsize{$-(25.6\pm3.5)$}	      &  \hspace{0.5cm} \scriptsize{$172.0$}                                          &\hspace{0.5cm} \scriptsize{$0.12$}	&\hspace{0.5cm}  \scriptsize{$1.4$}\\
	                           &\hspace{0.5cm} \scriptsize{$6.8\pm1.6$}	      &  \hspace{0.5cm} \scriptsize{$$}                                          &\hspace{0.5cm} \scriptsize{$$}	&\hspace{0.5cm}  \scriptsize{}\\ \cdashline{1-5}
\scriptsize{d}   	          &\hspace{0.5cm} \scriptsize{$-(16.3\pm3.4)$}	      &  \hspace{0.5cm} \scriptsize{$208.0$}                                          &\hspace{0.5cm} \scriptsize{$0.14$}	&\hspace{0.5cm}  \scriptsize{$0.1$}\\
	                           &\hspace{0.5cm} \scriptsize{$16.4\pm5.9$}	      &  \hspace{0.5cm} \scriptsize{$$}                                          &\hspace{0.5cm} \scriptsize{$$}	&\hspace{0.5cm}  \scriptsize{}\\ \cdashline{1-5}
\scriptsize{e}   	          &\hspace{0.5cm} \scriptsize{$-(19.7\pm3.2)$}	      &  \hspace{0.5cm} \scriptsize{$253.0$}                                          &\hspace{0.5cm} \scriptsize{$0.17$}	&\hspace{0.5cm}  \scriptsize{$0.1$}\\
	                           &\hspace{0.5cm} \scriptsize{$9.1\pm7.1$}	      &  \hspace{0.5cm} \scriptsize{$$}                                          &\hspace{0.5cm} \scriptsize{$$}	&\hspace{0.5cm}  \scriptsize{}\\ \cdashline{1-5}
\scriptsize{f}   	          &\hspace{0.5cm} \scriptsize{$-(11.7\pm9.1)$}	      &  \hspace{0.5cm} \scriptsize{$364.0$}                                          &\hspace{0.5cm} \scriptsize{$0.12$}	&\hspace{0.5cm}  \scriptsize{$0.2$}\\
	                           &\hspace{0.5cm} \scriptsize{$3.3\pm2.1$}	      &  \hspace{0.5cm} \scriptsize{$$}                                          &\hspace{0.5cm} \scriptsize{$$}	&\hspace{0.5cm}  \scriptsize{}\\ \cdashline{1-5}
\scriptsize{g}   	          &\hspace{0.5cm} \scriptsize{$-(8.9\pm5.2)$}	      &  \hspace{0.5cm} \scriptsize{$401.0$}                                          &\hspace{0.5cm} \scriptsize{$0.16$}	&\hspace{0.5cm}  \scriptsize{$0.2$}\\
	                           &\hspace{0.5cm} \scriptsize{$5.3\pm2.4$}	      &  \hspace{0.5cm} \scriptsize{$$}                                          &\hspace{0.5cm} \scriptsize{$$}	&\hspace{0.5cm}  \scriptsize{}\\ \cdashline{1-5}
\scriptsize{h}   	          &\hspace{0.5cm} \scriptsize{$-(7.9.\pm6.5)$}	      &  \hspace{0.5cm} \scriptsize{$508.0$}                                          &\hspace{0.5cm} \scriptsize{$0.22$}	&\hspace{0.5cm}  \scriptsize{$0.6$}\\
	                           &\hspace{0.5cm} \scriptsize{$5.8\pm2.7$}	      &  \hspace{0.5cm} \scriptsize{$$}                                          &\hspace{0.5cm} \scriptsize{$$}	&\hspace{0.5cm}  \scriptsize{}\\ \cdashline{1-5}
\scriptsize{i}   	          &\hspace{0.5cm} \scriptsize{$-(19.9\pm3.8)$}	      &  \hspace{0.5cm} \scriptsize{$840.0$}                                          &\hspace{0.5cm} \scriptsize{$0.11$}	&\hspace{0.5cm}  \scriptsize{$0.5$}\\
	                           &\hspace{0.5cm} \scriptsize{$10.2\pm2.5$}	      &  \hspace{0.5cm} \scriptsize{$$}                                          &\hspace{0.5cm} \scriptsize{$$}	&\hspace{0.5cm}  \scriptsize{}\\ \cdashline{1-5}

\hline
\end{tabular}
\end{table}
%


\begin{table}[]
\centering \renewcommand{\arraystretch}{1.7}\addtolength{\tabcolsep}{-2pt}
\caption{The best-fit parameters of the XRT  spectrum using a PL  and a PL plus a BB model. The first line in each time window corresponds to the best-fit parameters with the PL and the second one with the PL plus the BB model.}
\begin{tabular}{cccccc}
\hline
\multicolumn{1}{c}{\begin{tabular}[c]{@{}c@{}}Time Window\\ (s)\end{tabular}} & \multicolumn{1}{c}{\begin{tabular}[c]{@{}c@{}}$n_{\rm H} \times 10^{22}$ \\ (${\rm cm^{-2}}$)\end{tabular}} &  \multicolumn{1}{c}{\begin{tabular}[c]{@{}c@{}}Spectral Index\\ $\beta$ \end{tabular}} &     \multicolumn{1}{c}{\begin{tabular}[c]{@{}c@{}}KT\\ (keV) \end{tabular}} &     \multicolumn{1}{c}{\begin{tabular}[c]{@{}c@{}}Reduced\\ $\chi^{2}$ \end{tabular}} &     \multicolumn{1}{c}{\begin{tabular}[c]{@{}c@{}}$f_{\rm BB}/f_{\rm BL}$\\ \end{tabular}}\\
\hline
\hline                                                                        
{100 - 150} & $0.37^{+0.12}_{-0.10}$  & $2.19^{+0.12}_{-0.10}$ & - & $1.19$ & - \\ 
 & $0.17^{+0.29}_{-0.17}$ & $2.10^{+1.80}_ {-1.19}$ & $0.65^{+0.16}_{-0.11}$ & $1.16$ & $0.8621$ \\ \hline
 
 {150 - 200} & $0.66^{+0.26}_{-0.20}$  & $2.29^{+0.35}_{-0.32}$ & - & $1.24$ & - \\
 & $0.08^{+0.08}_{-0.08}$ & $-3.00^{+err}_{-err}$ & $0.67^{+0.07}_{-0.04}$ & $1.20$ & $2.52$ \\  \hline

{200 - 230} & $1.26^{+0.67}_{ -0.49}$  & $2.46^{+0.60}_{-0.53}$ & - & $0.76$ & - \\ 
 & $3.32^{+2.2}_{-1.5}$ & $-3.39 ^{+1.06}_{-0.46}$ & $0.07^{+0.01}_{-0.02}$ & $0.75$ & $389.27$ \\   \hline

{250 - 299} & $1.97^{+1.21}_{-0.92}$  & $3.02^{+1.13}_{-0.97}$ & - & $0.95$ & - \\ 
 & $3.49^{+3.05}_{-2.55}$ & $5.61 ^{+err}_{-err}$ & $1.76^{+1.24}_{-0.95}$ & $0.96$ & $0.001$ \\   \hline

{300 - 499} & $1.25^{+0.38}_{-0.32}$  & $2.64^{+0.44}_{-0.41}$ & - & $1.13$ & - \\ 
 & $1.40^{+1.32}_{-0.92}$ & $3.63 ^{+3.03}_{-3.38}$ & $0.86^{+95.5}_{-0.64}$ & $1.16$ & $0.07$ \\   \hline

{500 - 700} & $1.00^{+0.35}_{-0.29}$  & $2.58^{+0.47}_{-0.44}$ & - & $0.80$ & - \\ 
 & $1.52^{+0.97}_{-0.92}$ & $4.69 ^{+1.95}_{-2.57}$ & $0.92^{+0.81}_{-0.34}$ & $0.76$ & $0.02$ \\   \hline

{700 - 900} & $1.37^{+0.43}_{-0.36}$  & $2.55^{+0.42}_{-0.39}$ & - & $1.36$ & - \\ 
 & $1.07^{+1.07}_{-0.77}$ & $2.32 ^{+2.83}_{-2.72}$ & $0.63^{+0.26}_{-0.19}$ & $1.34$ & $0.18$ \\   \hline

{1000 - 1200} & $0.91^{+0.14}_{-0.13}$  & $2.05^{+0.16}_{-0.15}$ & - & $1.15$ & - \\ 
 & $0.91^{+0.31}_{-0.30}$ & $2.17 ^{+0.64}_{-1.03}$ & $1.16^{+0.91 }_{-0.24}$ & $1.15$ & $0.01$ \\   \hline

{1200 - 1299} & $0.79^{+0.12}_{-0.12}$  & $1.87^{+0.14}_{-0.13}$ & - & $1.16$ & - \\
 & $0.94^{+0.18}_{-0.16}$ & $1.98 ^{+0.16}_{-0.15}$ & $0.05^{+0.01}_{-0.01}$ & $1.14$ & $14.76$ \\  \hline

{1300 - 1399} & $0.98^{+0.13}_{-0.12}$  & $2.08^{+0.14}_{-0.13}$ & - & $1.19$ & - \\
 & $0.49^{+0.33}_{-0.24}$ & $1.29 ^{+0.66}_{-1.31}$ & $0.63^{+0.11}_{-0.08}$ & $1.19$ & $0.53$ \\   \hline

{1400 - 1500} & $1.09^{+0.15}_{-0.14}$  & $2.21^{+0.15}_{-0.14}$ & - & $1.24$ & - \\
 & $0.43^{+0.32}_{-0.18}$ & $1.05 ^{+0.79}_{-1.59}$ & $0.64^{+0.07}_{-0.07}$ & $1.19$ & $0.91$ \\ \hline

{1500 - 1599} & $0.93^{+0.13}_{-0.11}$  & $2.05^{+0.13}_{-0.13}$ & - & $1.44$ & - \\
 & $0.73^{+0.26}_{-0.25}$ & $1.71 ^{+0.56}_{-0.60}$ & $0.51^{+0.28}_{-0.24}$ & $1.44$ & $0.16$ \\   \hline

{1600 - 1700} & $1.20^{+0.15}_{-0.14}$  & $2.20^{+0.14}_{-0.14}$ & - & $1.26$ & - \\ 
 & $0.88^{+0.33}_{-0.34}$ & $1.79 ^{+0.48}_{-0.87}$ & $0.59^{+0.37}_{-0.11}$ & $1.30$ & $0.23$ \\

\hline                                                                              
\end{tabular}
\label{tab:my-table}
\end{table}

\begin{table}
\centering \renewcommand{\arraystretch}{2}\addtolength{\tabcolsep}{3pt}
\caption{The best-fit parameters found from the multi-wavelength observations.} \label{table:parameters}
\begin{tabular}{c c c }
 \hline \hline
\scriptsize{Parameter}   &\hspace{0.5cm}   \scriptsize{Values}                                              \\ 
\hline \hline
Spin-down Magnetar\\

\scriptsize{B ($\times 10^{16}\,{\rm G}$)}   	          &\hspace{0.5cm} \scriptsize{$ 3.0\pm0.3 $}	                                         \\
\scriptsize{P ($10^{-3}\,{\rm s}$)}   	          &\hspace{0.5cm} \scriptsize{$ 1.2\pm 0.1 $}	                                        \\
\scriptsize{$t_{\rm fb}\,({\rm s})$}   	          &\hspace{0.5cm} \scriptsize{$ (1.5\pm0.2)\times 10^4 $}	                                         \\ 
\hline
Synchrotron FS model\\

\scriptsize{$\varepsilon_{\rm B}$ ($10^{-4}$)}   	          &\hspace{0.5cm} \scriptsize{$ 1.1\pm0.1 $}	                                             \\
\scriptsize{$\varepsilon_{\rm e}$ ($10^{-1}$)}   	          &\hspace{0.5cm} \scriptsize{$ 0.8\pm0.1 $}	                                         \\
\scriptsize{${\rm p}$}   	                                                    &\hspace{0.5cm} \scriptsize{$ 2.3\pm0.2 $}	                                       \\
\scriptsize{${\rm n}$ (${10^{-1}}\,{\rm cm^{-3}}$)}   	                           &\hspace{0.5cm} \scriptsize{$ 1.0\pm0.1 $}	                                         \\
\scriptsize{${\rm E}$ (${10^{51}}\,{\rm erg}$)}   	                           &\hspace{0.5cm} \scriptsize{$ 2.4\pm0.2 $}	                                         \\
\hline \hline
\end{tabular}
\end{table}

\clearpage

\clearpage
\clearpage
\clearpage
\end{document}